\documentclass[final,3p,times,twocolumn]{elsarticle}

\usepackage{geometry}
\usepackage[utf8]{inputenc}
\usepackage{graphicx}
\usepackage{float}
\usepackage{caption}
\usepackage{subcaption}

\usepackage[T1]{fontenc}
\usepackage[american]{babel}
\usepackage{fix-cm}
\usepackage{textcomp}
\usepackage{helvet}
\usepackage{float}
\usepackage{mathtools}
\usepackage{graphicx}
\usepackage{newfloat}
\usepackage{amssymb}
\usepackage[switch]{lineno}
\usepackage{epigraph}
\usepackage{parskip}
\usepackage{bm}
\usepackage{enumitem}
\usepackage{rotating}
\usepackage[svgnames]{xcolor}
\makeatletter
\newcommand\flausr{\@fleqntrue}
\makeatother

\usepackage{hyperref}
\usepackage{caption}
\DeclareCaptionType{InfoBox}

\usepackage{color}

\usepackage{colortbl}
\definecolor{hellgrau}{rgb}{0.9,0.9,0.9}
\definecolor{mygreen}{RGB}{28,172,0} 
\definecolor{mylilas}{RGB}{170,55,241}

\usepackage{amsmath}
\usepackage{amsfonts}
\usepackage{tikz}
\usepackage{tabularx}
\usepackage{svg}
\usepackage{array}
\usepackage{longtable}
\usepackage{eqnarray}
\usepackage{pdflscape}
\usepackage{comment}
\usepackage{listings}
\usepackage{multirow}
\usepackage{units}
\usepackage{booktabs}
\usepackage{varwidth}
\usepackage{algpseudocode}
\usepackage{algorithm}
\usepackage{placeins}

\usepackage{url}
\hypersetup{
	colorlinks=true,
	linkcolor=blue,
	filecolor=magenta,
	urlcolor=blue,
	citecolor=blue,
}
\usepackage{cleveref}
\graphicspath{ {./figures/} }
\def\BibTeX{{\rm B\kern-.05em{\sc i\kern-.025em b}\kern-.08em
		T\kern-.1667em\lower.7ex\hbox{E}\kern-.125emX}}

\makeatletter
\renewcommand{\fnum@figure}{Fig. \thefigure}
\makeatother

\newcommand*{\figref}[2][]{%
	\hyperref[{#2}]{%
		Fig.~\ref*{#2}%
		\ifx\\#1\\%
		\else
		\,#1%
		\fi
	}%
}

\newcommand*{\tabref}[2][]{%
	\hyperref[{#2}]{%
		Table~\ref*{#2}%
		\ifx\\#1\\%
		\else
		\,#1%
		\fi
	}%
}

\newtheorem{thm}{Theorem}
\newtheorem{lem}[thm]{Lemma}
\newdefinition{rmk}{Remark}
\newproof{pf}{Proof}
\newproof{pot}{Proof of Theorem \ref{thm2}}

\newcommand{\blueline}{\raisebox{2pt}{\tikz{\draw[-,cyan!35!blue,solid,line width = 1pt](0,0) -- (3mm,0);}}}
\newcommand{\orangeline}{\raisebox{2pt}{\tikz{\draw[-,yellow!30!orange,solid,line width = 0.9pt](0,0) -- (3mm,0);}}}
\newcommand{\redlinedashed}{\raisebox{2pt}{\tikz{\draw[-,red!100!orange,dashed,line width = 0.9pt](0,0) -- (3mm,0);}}}

\newcommand{\redline}{\raisebox{2pt}{\tikz{\draw[-,red!100!orange,solid,line width = 1pt](0,0) -- (3mm,0);}}}

\newcommand{\orangelinedashed}{\raisebox{2pt}{\tikz{\draw[-,yellow!20!orange,dashed,line width = 0.8pt](0,0) -- (3mm,0);}}}

\newcommand{\greenlinedashed}{\raisebox{2pt}{\tikz{\draw[-,black!35!green,dashed,line width = 0.8pt](0,0) -- (3mm,0);}}}
\newcommand{\brownlinedashed}{\raisebox{2pt}{\tikz{\draw[-,black!0!brown,dashed,line width = 0.8pt](0,0) -- (3mm,0);}}}
\newcommand{\purplelinedashed}{\raisebox{2pt}{\tikz{\draw[-,violet!40!violet,dashed,line width = 0.8pt](0,0) -- (3mm,0);}}}
\newcommand{\purpleline}{\raisebox{2pt}{\tikz{\draw[-,violet!40!violet,solid,line width = 1pt](0,0) -- (3mm,0);}}}
\newcommand{\greenline}{\raisebox{2pt}{\tikz{\draw[-,black!35!green,solid,line width = 1pt](0,0) -- (3mm,0);}}}

\newcommand{\cyanline}{\raisebox{2pt}{\tikz{\draw[-,cyan!30!cyan,solid,line width = 1pt](0,0) -- (3mm,0);}}}

\definecolor{orange}{rgb}{0.85, 0.55, 0}

\makeatletter
\def\ps@pprintTitle{%
	\let\@oddhead\@empty
	\let\@evenhead\@empty
	\let\@oddfoot\@empty
	\let\@evenfoot\@oddfoot
}
\makeatother

\journal{Elsevier}
\title{An environmental disturbance observer framework for autonomous surface vessels} 

\begin{document} 

\begin{frontmatter}
	
	\author{Daniel~Menges\corref{cor1}\fnref{address1}}
	\ead{daniel.menges@ntnu.no}
	\cortext[cor1]{Corresponding author}
	\author[address1,address2]{Adil Rasheed}
	\ead{adil.rasheed@ntnu.no}
	
	\address[address1]{Department of Engineering Cybernetics, Norwegian University of Science and Technology}
        \address[address2]{Mathematics and Cybernetics, SINTEF Digital}

	\begin{abstract}
		This paper proposes a robust disturbance observer framework for maritime autonomous surface vessels considering model and measurement uncertainties. The core contribution lies in a nonlinear disturbance observer, reconstructing the forces on a vessel impacted by the environment. For this purpose, mappings are found leading to synchronized global exponentially stable error dynamics. With the stability theory of Lyapunov, it is proven that the error converges exponentially into a ball, even if the disturbances are highly dynamic. Since measurements are affected by noise and physical models can be erroneous, an unscented Kalman filter (UKF) is used to generate more reliable state estimations. In addition, a noise estimator is introduced, which approximates the noise strength. Depending on the severity of the measurement noise, the observed disturbances are filtered through a cascaded structure consisting of a weighted moving average (WMA) filter, a UKF, and the proposed disturbance observer. To investigate the capability of this observer framework, the environmental disturbances are simulated dynamically under consideration of different model and measurement uncertainties. It can be seen that the observer framework can approximate dynamical forces on a vessel impacted by the environment despite using a low measurement sampling rate, an erroneous model, and noisy measurements.
	\end{abstract}
	
\end{frontmatter}
\pagenumbering{arabic}
\section{Introduction}
In the existing literature, several control concepts have been proposed for path following and collision avoidance for safe navigation. In this context, a precise perception of the environment is one of the essential elements of the vessel's situational awareness for safe sea operations. For example, sea currents might drift the vessel away from the desired path, while the wind can affect the vessel in its heading by creating a torque on the vessel's hull. If control algorithms are not considering these influences \cite{Do2005}, safe and comfortable sea operations are not guaranteed. To accommodate these factors, some previous works consider the environmental impact. For instance, a disturbance observer-based control for the dynamic positioning of vessels is proposed in \cite{Wei2022}. The proposal assumes a constant damping matrix and considers weak disturbances, such as in \cite{Huang2015}, where a global stable tracking control of under-actuated ships with input saturation is developed. In \cite{Hu2022}, a robust synchronization for under-actuated vessels based on a disturbance observer is introduced, where slowly varying disturbances are considered. The results showcase that the disturbance estimations adapt inertly to the actual disturbances. In addition, it is mentioned that the existing schemes require exact velocity measurements. Since measurement uncertainties are inevitable, this is an essential issue that has to be addressed. Measurement noise propagates through disturbance observers leading to inadvertent noisy results. An overview of recent advances in coordinated control of multiple autonomous surface vessels is given in \cite{Peng2021a}, where it is mentioned that the central focus of current investigations is guaranteeing the stability and robustness of motion control laws in the presence of uncertainties and disturbances. 
In addition, sliding mode control approaches based on disturbance observers are proposed \cite{Li2015, Chen2019}. The results of the observer showcased in \cite{Chen2019} demonstrate that the disturbances inertly adapt with various adaptation speeds concerning the dedicated system states. 
In a recent study by \cite{Selvaraj2020}, an approach to reject uncertainty and disturbance in complex dynamical networks using truncated predictive control is discussed. The study guarantees a robust synchronization of the networks by incorporating input delay and utilizing the Lyapunov stability theory. However, it is important to note that the simulations conducted do not account for the potential impact of measurement noise.
Besides, a global asymptotic regulation control for MIMO mechanical systems with unknown model parameters and disturbances is proposed \cite{Hu2019}. As in many other proposals, a constant damping matrix is assumed, and the Coriolis matrix is completely neglected. One of the few disturbance observers presented in \cite{Bouteraa2022} for tracking control considers model uncertainties. However, measurement uncertainties are neglected, and similar to the previously listed work, the damping matrix is oversimplified, making its use in real applications questionable.
A backstepping control approach based on a disturbance observer and a neural network is presented in \cite{Duong2022}, where the vessel stability concerning the roll motion is studied. Here, as well as in \cite{Xu2022}, only slowly varying disturbances are considered, while it is assumed that the measurements are entirely reliable. Despite ignoring model and measurement uncertainties, a widely used disturbance observer is presented in \cite{Do2010}. The proposal does not provide insight into the error dynamics of why it is difficult to guarantee sufficient adaptation speed regarding all observed disturbances. For this reason, synchronizing the error dynamics is a convenient way to supply controllers with consistent estimations. A detailed survey about the most popular disturbance observers is given in \cite{Gu2022}.  
More recently, with the advancement of data-driven modeling, machine learning approaches for observing disturbances are coming up. A disturbance observer constructed by fusion of neural networks and minimal learning parameterization is proposed in \cite{Huang2019} for robust dynamic positioning control under consideration of model and servo-system uncertainties, while \cite{Qiang2019} proposes adaptive neural network auto-berthing control of marine ships. The latter mentions that external disturbances can not be approximated by neural networks. In \cite{Peng2021}, a data-driven adaptive disturbance observer for model-free trajectory tracking control of maritime autonomous surface vessels is introduced. Common to all the data-driven approaches is that the models are trained with historical data and have a memory stack for online learning. Once again, most of such works are limited to scenarios with very weak disturbances. Since data-driven models are prone to overfitting and suffer from poor generalization outside the training scenarios, they are unfit to handle previously unseen severe and dynamic disturbances \cite{Trivedi2021}. 
Furthermore, the impact and rejection of disturbances and uncertainties are studied in several other applications, including autonomous aircraft \cite{Emami2019}, unmanned driving robotic vehicles \cite{Chen2019a}, and in general nonholonomic mobile robots \cite{Kim2003}.
The estimation of environmental disturbances is relevant for trajectory tracking, such as described in \cite{Emami2019}, where an intelligent trajectory tracking of an aircraft in the presence of internal and external disturbances is discussed using neural network-based model predictive control. While the impact of model uncertainties is considered in the regarded proposals, the influence of measurement uncertainties is completely ignored.

To summarize, none of the proposed approaches demonstrate robust behavior under several environmental conditions by considering measurement uncertainties. Moreover, most proposals disregard that measurements must be treated discretely since the sampling frequency of measurements is limited. Hence, this work tries to handle the previously described weaknesses in a novel observer framework by avoiding model simplifications but considering model and measurement uncertainties. To this end, the current work attempts to answer the following questions:

\begin{itemize}
	\item Can we reconstruct the unknown disturbances despite measurement uncertainties? 
	\item Is it possible to reconstruct the disturbances despite using an unreliable model?
	\item Is it possible to observe the disturbances in situations where the sampling rate of the measurements is very low? 
	\item Can a disturbance observer be designed where we can synchronize the adaptation speed of all observed disturbances?
\end{itemize}
To the best of our knowledge, none of the previous works have addressed all the research questions mentioned in a single study. 

For a better comprehension of the work presented, the relevant theory, including the original contribution, is presented in Section \ref{sec:theory}. Section \ref{sec:methodandsetup} presents all the details required to reproduce the results presented in the article. Results and their discussions are presented in Section \ref{section: Results} and finally, Section \ref{sec:conclusionandfuturework} concludes the current work.  

\section{Theory}
\label{sec:theory}
In this section, we present a brief overview of the theory that is required for a better comprehension of the work presented. We begin by explaining the model of the vessel used in Section \ref{section: Model} followed by introducing the weighted moving average in Section \ref{section: WMA}, and a description of the unscented Kalman filter in Section \ref{section: UKF}. The original theoretical contribution of the work is presented in Sections \ref{section: Observerdesign} and \ref{section: Framework}.

\subsection{Model}\label{section: Model}
\begin{figure}[ht]
	\centering
	\includegraphics[width=\linewidth]{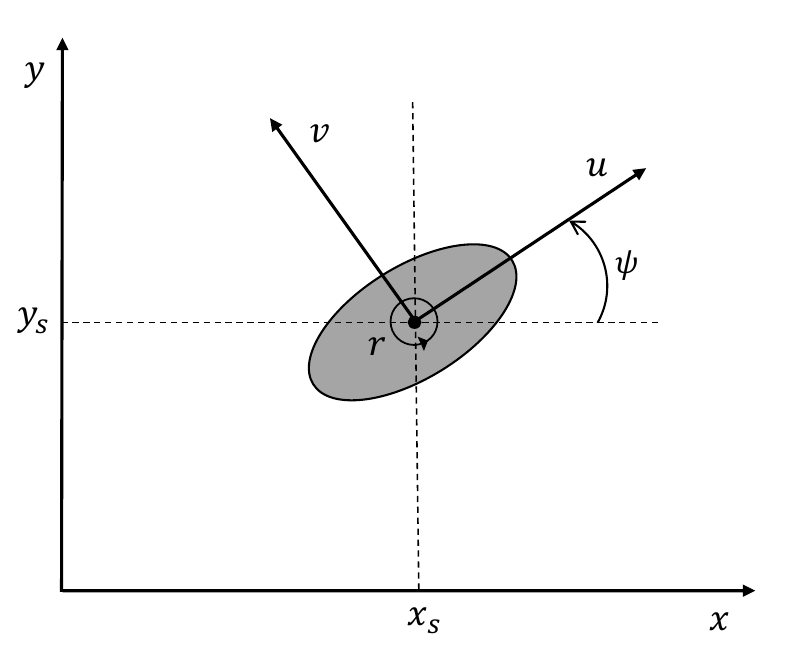}
	\caption{Kinematics of a vessel}
	\label{fig:Ship_kinematics}
\end{figure}

Considering the vessel's coordinates $\boldsymbol{\eta} = [x_s,  y_s,  \psi]^\top$, where $x_s$ and $y_s$ describes the vessel's position with regard to the global coordinates $[x, y]^\top$, and $\psi$ is the vessel's heading. Furthermore, the velocities are denoted as $\boldsymbol{\nu} = [u,  v,  r]^\top$, where $u$ is the surge, $v$ describes the sway, and $r$ characterizes the rotational speed regarding yaw. With these relations, the kinematics of a vessel can be expressed by
\begin{linenomath*}
	\begin{equation}
		\boldsymbol{\dot{\eta}}=\boldsymbol{R}_{\mathrm{rot}}(\psi)\boldsymbol{\nu},
	\end{equation}
\end{linenomath*}
where $\boldsymbol{R}_{\mathrm{rot}}(\psi)$ denotes the rotational matrix, given by
\begin{linenomath*}
	\begin{equation}
		\boldsymbol{R}_{\mathrm{rot}}(\psi)=\begin{bmatrix}
			\mathrm{cos}(\psi) && -\mathrm{sin}(\psi) && 0\\
			\mathrm{sin}(\psi) && \mathrm{cos}(\psi) && 0 \\
			0 && 0 && 1
		\end{bmatrix}.
	\end{equation}
\end{linenomath*}
\figref{fig:Ship_kinematics} depicts the relations between the different vessel states.

The dynamics of a surface vessel adopted from \cite{Fossen2011} can be expressed as
\begin{linenomath*}
	\begin{equation}
		\boldsymbol{M}\boldsymbol{\dot{\nu}}+\boldsymbol{D}(\boldsymbol{\nu})\boldsymbol{\nu}+\boldsymbol{C}(\boldsymbol{\nu})\boldsymbol{\nu}=\boldsymbol{\tau}+\boldsymbol{\tau_d}, \label{eq: dynamics}
	\end{equation}
\end{linenomath*}
where $\boldsymbol{M}$ denotes the mass matrix, $\boldsymbol{D}(\boldsymbol{\nu})$ characterizes the nonlinear damping matrix, $\boldsymbol{C}(\boldsymbol{\nu})$ describes the Coriolis matrix, $\boldsymbol{\tau}$ is the control input, and $\boldsymbol{\tau_d}$ are all the environmental disturbances. Therefore,
\begin{linenomath*}
	\begin{equation}
		\boldsymbol{\tau_d}=\boldsymbol{\tau}_{\mathrm{wind}}+\boldsymbol{\tau}_{\mathrm{wave}}+\boldsymbol{\tau}_{\mathrm{current}},
	\end{equation}
\end{linenomath*}
which implies that the disturbances are specified as the sum of forces induced by wind, waves, and sea currents.
Rewriting the dynamical expression in state space representation leads to
\begin{linenomath*}
	\begin{equation}
		\boldsymbol{\dot{\nu}} = \boldsymbol{M}^{-1}\left(-\boldsymbol{D}(\boldsymbol{\nu})\boldsymbol{\nu}-\boldsymbol{C}(\boldsymbol{\nu})\boldsymbol{\nu}+\boldsymbol{\tau}+\boldsymbol{\tau_d}\right). \label{model_equation}
	\end{equation}
\end{linenomath*}

Considering a body-fixed coordinate system $[x_b, y_b]^\top$ with center $[x_s, y_s]^\top$, where $x_b$ is directed to surge and $y_b$ is directed to sway. Under the assumption that the rigid body has a homogeneous mass distribution and is symmetric concerning the $x_b y_b$-plane, the mass matrix is given by
\begin{linenomath*}
	\begin{equation}
		\boldsymbol{M}_{\mathrm{ideal}} = \begin{bmatrix}
			m && 0 && 0\\
			0 && m && mx_g\\
			0 && mx_g && I_z
		\end{bmatrix}, \label{eq:M_ideal}
	\end{equation}
\end{linenomath*}
where $m$ defines the mass of the rigid body, $x_g$ is the center of gravity regarding the $x_b$-axis, and $I_z$ characterizes the moment of inertia concerning yaw. 
Since expression \eqref{eq:M_ideal} requires simplifications leading to model uncertainties, we introduce a generalized, symmetric, positive definite mass matrix 
\begin{linenomath*}
	\begin{equation}
		\boldsymbol{M} = \begin{bmatrix}
			m_{11} && 0 && 0\\
			0 && m_{22} && m_{23}\\
			0 && m_{32} && m_{33}
		\end{bmatrix},
	\end{equation}
\end{linenomath*}
such that five instead of three unknown parameters must be identified. The parameters are assumed to be constant since the mass matrix is not dependent on the system states $\boldsymbol{\nu}$.
Based on later derivations, the entries of the inverted mass matrix are denoted as
\begin{linenomath*}
	\begin{equation}
		\boldsymbol{M}^{-1} = \begin{bmatrix}
			\kappa_{11} && 0 && 0\\
			0 && \kappa_{22} && \kappa_{23}\\
			0 && \kappa_{32} && \kappa_{33}
		\end{bmatrix}.
	\end{equation}
\end{linenomath*}

\begin{lem}
	If the matrix $\boldsymbol{M}$ is symmetric and invertible, its inverse $\boldsymbol{M}^{-1}$ is likewise symmetric, since $(\boldsymbol{M}^{-1})^\top =~(\boldsymbol{M}^\top)^{-1}$. Since the entries of the mass matrix are positive, $\boldsymbol{M}$ is positive definite and thus invertible.
\end{lem}

Moreover, it is assumed that the damping matrix is symmetric, and the Coriolis matrix is skew-symmetric such as in \cite{Fossen2011}, given by
\begin{linenomath*}
	\begin{align}
		\boldsymbol{D}(\boldsymbol{\nu}) &= \begin{bmatrix}
			d_{11}(\boldsymbol{\nu}) && 0 && 0\\
			0 && d_{22}(\boldsymbol{\nu}) && d_{23}(\boldsymbol{\nu})\\
			0 && d_{32}(\boldsymbol{\nu}) && d_{33}(\boldsymbol{\nu})
		\end{bmatrix},\\ 
		\boldsymbol{C}(\boldsymbol{\nu}) &= \begin{bmatrix}
			0 && 0 && c_{13}(\boldsymbol{\nu})\\
			0 && 0 && c_{23}(\boldsymbol{\nu})\\
			c_{31}(\boldsymbol{\nu}) && c_{32}(\boldsymbol{\nu}) && 0
		\end{bmatrix}.
	\end{align}
\end{linenomath*}
The entries of the matrices are defined as
\begin{linenomath*}
	\begin{subequations}
		\begin{align}
			d_{11}(\boldsymbol{\nu}) &= -X_u-X_{|u|u}|u|-X_{uuu}u^2,\\
			d_{22}(\boldsymbol{\nu}) &= -Y_v-Y_{|v|v}|v|-Y_{|r|v}|r|-Y_{vvv}v^2,\\
			d_{23}(\boldsymbol{\nu}) &= -Y_r-Y_{|v|r}|v|-Y_{|r|r}|r|,\\
			d_{32}(\boldsymbol{\nu}) &= -N_v-N_{|v|v}|v|-N_{|r|v}|r|,\\
			d_{33}(\boldsymbol{\nu}) &= -N_r-N_{|v|r}|v|-N_{|r|r}|r|-N_{rrr}r^2,
		\end{align}
	\end{subequations}
\end{linenomath*}
and
\begin{linenomath*}
	\begin{subequations}
		\begin{align}
			c_{13}(\boldsymbol{\nu}) &= -m_{22}v-m_{23}r,\\
			c_{23}(\boldsymbol{\nu}) &= m_{11}u,\\
			c_{31}(\boldsymbol{\nu}) &= -c_{13}(\boldsymbol{\nu}),\\
			c_{32}(\boldsymbol{\nu}) &= -c_{23}(\boldsymbol{\nu}),
		\end{align}
	\end{subequations}
\end{linenomath*}
causing highly nonlinear model dynamics, where $X_u$, $X_{|u|u}$, $X_{uuu}$, $Y_v$, $Y_{|v|v}$, $Y_{|r|v}$, $Y_{vvv}$, $Y_r$, $Y_{|v|r}$, $Y_{|r|r}$, $N_v$, $N_{|v|v}$, $N_{|r|v}$, $N_r$, $N_{|v|r}$, $N_{|r|r}$, and $N_{rrr}$ are hydrodynamic parameters.

\subsection{Weighted Moving Average}\label{section: WMA}
Moving averages are used in fields such as statistics, soft computing, economics, operational research, and engineering \cite{Merigo2019}. The simple moving average (SMA) at time step $k$ with a window size $w$ is defined by
\begin{align}
	y_{SMA,k} = \frac{1}{w}\cdot\sum_{i=1}^{w} y_{k-w+i}, \label{eq:WMA}
\end{align}
where $y_k$ is the measurement at time step $k$.
Moving averages are popular tools for time-series smoothing \cite{Yager2008}. However, the disadvantage of the simple moving average (SMA) is its relatively strong delay depending on the window size $w$. Considering measurement noise, strong smoothing properties with little delay are desired.
The weighted moving average WMA is similar to the SMA but with a weighting of the past samples depending on their recency. Hence, the WMA at time step $k$ with a window size $w$ is expressed by
\begin{align}
	y_{WMA,k} = \frac{1}{\sum_{i=1}^{w}i}\cdot\sum_{i=1}^{w} i\cdot y_{k-w+i}. \label{eq:WMA}
\end{align}
The WMA is particularly beneficial if the samples are dynamic since it approximates a smoothed version of the measurements with little delay. As a result, the WMA is more sensitive to strong noise while staying in the actual range of the measurements.

\subsection{Unscented Kalman Filter}\label{section: UKF}
The unscented Kalman filter (UKF), proposed in \cite{Julier1997}, is proved to be a very powerful tool for state estimation of nonlinear systems with measurement noise and model uncertainty. Considering the time-discrete nonlinear dynamical system
\begin{linenomath*}
	\begin{align}
		\boldsymbol{\mathrm{x}}_{k+1} &= \boldsymbol{f}(\boldsymbol{\mathrm{x}}_{k},\boldsymbol{\mathrm{w}}_{k}) \label{UKF_model},\\
		\boldsymbol{\mathrm{y}}_{k} &= \boldsymbol{g}(\boldsymbol{\mathrm{x}}_{k},\boldsymbol{\mathrm{n}}_{k}) \label{UKF_measurement},
	\end{align}
\end{linenomath*}
where $\boldsymbol{f}$ and $\boldsymbol{g}$ are non-linear functions, $\boldsymbol{\mathrm{x}}_{k}$ describes the model states, and $\boldsymbol{\mathrm{y}}_{k}$ characterizes the measurement model at time step $k$.
The model and measurement uncertainties are modeled as Gaussian noise with zero mean. Hence, they are described by
\begin{linenomath*}
	\begin{align}
		\boldsymbol{\mathrm{w}}_{k}\sim\mathcal{N}(\boldsymbol{0},\boldsymbol{Q}_k), \label{model_noise}\\
		\boldsymbol{\mathrm{n}}_{k}\sim\mathcal{N}(\boldsymbol{0},\boldsymbol{R}_k), \label{meas_noise}
	\end{align}
\end{linenomath*}
where $\boldsymbol{Q}_k$ and $\boldsymbol{R}_k$ specify the covariance matrix of the model and measurement uncertainties, respectively. 
\\
The main idea of the UKF is to use sigma points distributed symmetrically in the area of the mean and gate them through the nonlinear functions. Assume the states $\boldsymbol{\mathrm{x}}\in \mathbb{R}^L$ have the estimated mean $\boldsymbol{\mathrm{\hat{x}}}$ and covariance $\boldsymbol{P}$, the sigma points of the entire sigma point matrix $\boldsymbol{\mathcal{X}}\in \mathbb{R}^{L\times (2L+1)}$ are generated by
\begin{linenomath*}
	\begin{subequations} 
		\begin{align}
			\boldsymbol{\mathcal{X}}_0 &= \boldsymbol{\mathrm{\hat{x}}}, \label{sigma_points1}\\
			\boldsymbol{\mathcal{X}}_i &= \boldsymbol{\mathrm{\hat{x}}} + \left(\sqrt{(L+\lambda)\boldsymbol{P}}\right)_i,\ \ \ \ \ \ i=1,...,L,\label{sigma_points2}\\
			\boldsymbol{\mathcal{X}}_i &= \boldsymbol{\mathrm{\hat{x}}} - \left(\sqrt{(L+\lambda)\boldsymbol{P}}\right)_{i-L},\ \ \ i=L+1,...,2L,\label{sigma_points3}
		\end{align}
	\end{subequations}
\end{linenomath*}
where $\lambda=\alpha^2(L+\kappa)-L$. Note that the bracket's index $i$ identifies the $i$-th column of the related matrix. The tuning parameter $\alpha$ describes the spread of the sigma points, usually set to $\alpha=10^{-3}$, and $\kappa$ is a secondary tuning parameter, usually set to $\kappa = 0$. The weights of each sigma point for calculating the means and covariances are defined as
\begin{linenomath*}
	\begin{align}
		&W^{m}_0 = \frac{\lambda}{L+\lambda},\\
		&W^{c}_0 = \frac{\lambda}{L+\lambda} + (1-\alpha^2+\beta),\\
		&W^{m}_i = W^{c}_i = \frac{1}{2(L+\lambda)}, \ \ \ i=1,...,2L,
	\end{align}
\end{linenomath*}
where $\beta$ is a tuning parameter, which is usually set to $\beta = 2$ for Gaussian distributions.

\textbf{Prediction step}
\\
Calculate $\boldsymbol{\mathcal{X}}$ according to \eqref{sigma_points1}-\eqref{sigma_points3}. 
Afterwards compute the following relations:
\begin{linenomath*}
	\begin{align}
		&\boldsymbol{\check{\mathrm{x}}} = \sum_{i=0}^{2L} W^{m}_i \boldsymbol{f}(\boldsymbol{\mathcal{X}}_i)\label{UKF_start}\\
		&\boldsymbol{\check{\mathrm{y}}} = \sum_{i=0}^{2L} W^{m}_i \boldsymbol{g}(\boldsymbol{\mathcal{X}}_i)\\
		&\boldsymbol{P}_x = \sum_{i=0}^{2L} W^{c}_i \left(\boldsymbol{f}(\boldsymbol{\mathcal{X}}_i)-\boldsymbol{\check{\mathrm{x}}} \right)\left(\boldsymbol{f}(\boldsymbol{\mathcal{X}}_i)-\boldsymbol{\check{\mathrm{x}}} \right)^\top\\
		&\boldsymbol{P}_y = \sum_{i=0}^{2L} W^{c}_i \left(\boldsymbol{g}(\boldsymbol{\mathcal{X}}_i)-\boldsymbol{\check{\mathrm{y}}} \right)\left(\boldsymbol{g}(\boldsymbol{\mathcal{X}}_i)-\boldsymbol{\check{\mathrm{y}}} \right)^\top\\
		&\boldsymbol{P}_{xy} = \sum_{i=0}^{2L} W^{c}_i \left(\boldsymbol{f}(\boldsymbol{\mathcal{X}}_i)-\boldsymbol{\check{\mathrm{x}}} \right)\left(\boldsymbol{g}(\boldsymbol{\mathcal{X}}_i)-\boldsymbol{\check{\mathrm{y}}} \right)^\top
	\end{align}
\end{linenomath*}

\textbf{Correction step}
\begin{linenomath*}
	\begin{align}
		&\boldsymbol{K} = \boldsymbol{P}_{xy}\boldsymbol{P}_y^{-1}\\ 
		&\boldsymbol{\mathrm{\hat{x}}} = \boldsymbol{\check{\mathrm{x}}}+\boldsymbol{K}(\boldsymbol{\mathrm{y}}-\boldsymbol{\check{\mathrm{y}}})\\
		&\boldsymbol{P} = \boldsymbol{P}_x-\boldsymbol{K}\boldsymbol{P}_y\boldsymbol{P}_x \label{UKF_end}
	\end{align}
\end{linenomath*}

Here, $\boldsymbol{\check{\mathrm{x}}}$ and $\boldsymbol{\check{\mathrm{y}}}$ are the predictions of the states and measurements, respectively, calculated by gating the sigma points $\boldsymbol{\mathcal{X}}$ through the nonlinear functions. The covariance matrices $\boldsymbol{P}_x$, $\boldsymbol{P}_y$, and the cross-covariance matrix $\boldsymbol{P}_{xy}$ are used for computing the updated covariance matrix $\boldsymbol{P}$ and the Kalman gain matrix $\boldsymbol{K}$. Hence, the Kalman gain matrix characterizes the trustworthiness of the predicted states $\boldsymbol{\check{\mathrm{x}}}$. If $\boldsymbol{K}$ has small values, the predictions are well-performing. Otherwise, the measurements $\boldsymbol{\mathrm{y}}$ are weighted stronger to correct the prediction inaccuracy.
As a result, the estimations $\boldsymbol{\mathrm{\hat{x}}}$ result in a smoothed version of the measurements $\boldsymbol{\mathrm{x}}_{m}$.  

\subsection{Disturbance Observer Design}\label{section: Observerdesign}
In the following, the original theoretical contribution of this article is explained. Two assumptions are made to formulate the disturbance observer.

\textbf{Assumptions:}
\begin{itemize}
	\item[1.] It is assumed that the vessel's velocities $\boldsymbol{\nu}$ are measured. However, the measurements are expected to have uncertainties, described in more detail in Section \ref{section: Framework}. 
	\item[2.] It is assumed that the disturbances change slowly such that $\boldsymbol{\dot{\tau}_d} \approx \boldsymbol{0}$ holds for short time intervals. This assumption is required for the observer design. Later, it is proven that the disturbances can also be highly dynamic.
\end{itemize}

To design the dynamical observer, we define the following relations
\begin{linenomath*}
	\begin{align}
		\boldsymbol{\hat{\tau}_d}&=\boldsymbol{\zeta} + \boldsymbol{\mu}(\boldsymbol{\nu}),\\
		\boldsymbol{\dot{\zeta}}&=\boldsymbol{h}(\boldsymbol{\nu},\boldsymbol{\hat{\tau}_d}),
	\end{align}
\end{linenomath*}
where the estimation of the disturbances $\boldsymbol{\hat{\tau}_d}$ is defined as the sum of an observer variable $\boldsymbol{\zeta}$ and an unknown mapping $\boldsymbol{\mu}(\boldsymbol{\nu})$. 
Hence, the error of the observer is defined as
\begin{linenomath*}
	\begin{equation}
		\boldsymbol{z} = \boldsymbol{\tau_d}-\boldsymbol{\zeta}-\boldsymbol{\mu}(\boldsymbol{\nu}). \label{error}
	\end{equation}
\end{linenomath*}
The dynamics of the observer variable are defined as an unknown mapping $\boldsymbol{h}(\boldsymbol{\nu},\boldsymbol{\hat{\tau}_d})$.
Therefore, the goal is to find the mappings $\boldsymbol{\mu}(\boldsymbol{\nu})$ and $\boldsymbol{h}(\boldsymbol{\nu},\boldsymbol{\hat{\tau}_d})$, such that the estimation $\boldsymbol{\hat{\tau}_d}$ has a globally, asymptotically stable equilibrium at $\boldsymbol{z}=\boldsymbol{0}$, and converges to the manifold
\begin{linenomath*}
	\begin{equation}
		\boldsymbol{\mathcal{M}}=\left\{(\boldsymbol{\nu},\boldsymbol{\zeta})\in \mathbb{R}^3\times\mathbb{R}^3: \boldsymbol{\tau_d}-\boldsymbol{\zeta} - \boldsymbol{\mu}(\boldsymbol{\nu}) = \boldsymbol{0}\right\}.
	\end{equation}
\end{linenomath*}

The error dynamics are obtained by the time derivative of \eqref{error}, yielding
\begin{linenomath*}
	\begin{equation}
		\boldsymbol{\dot{z}} = \underbrace{\boldsymbol{\dot{\tau}_d}}_{\approx\boldsymbol{0}}-\boldsymbol{\dot{\zeta}}-\frac{\partial \boldsymbol{\mu}}{\partial \boldsymbol{\nu}}\dot{\boldsymbol{\nu}},
	\end{equation}
\end{linenomath*}
which leads to  
\begin{linenomath*}
	\begin{equation}
		\boldsymbol{\dot{z}}= -\boldsymbol{h}(\boldsymbol{\nu},\boldsymbol{\hat{\tau}_d})-\frac{\partial \boldsymbol{\mu}}{\partial \boldsymbol{\nu}}\boldsymbol{M}^{-1}\left(\left(-\boldsymbol{D}(\boldsymbol{\nu})-\boldsymbol{C}(\boldsymbol{\nu})\right) + \boldsymbol{\tau} + \boldsymbol{\tau_d} \right).\label{error_dynamics}
	\end{equation}
\end{linenomath*}

\begin{rmk}
	Since the error dynamics are described by \eqref{error_dynamics}, $\boldsymbol{\mu}(\boldsymbol{\nu})$ must be continuously differentiable with respect to $\boldsymbol{\nu}$.
\end{rmk}

To find suitable error dynamics, we define $\boldsymbol{h}(\boldsymbol{\nu},\boldsymbol{\hat{\tau}_d})$ as
\begin{linenomath*}
	\begin{equation}
		\boldsymbol{h}(\boldsymbol{\nu},\boldsymbol{\hat{\tau}_d}) =-\frac{\partial \boldsymbol{\mu}}{\partial \boldsymbol{\nu}}\boldsymbol{M}^{-1}\left(\left(-\boldsymbol{D}(\boldsymbol{\nu})-\boldsymbol{C}(\boldsymbol{\nu})\right) + \boldsymbol{\tau} + \boldsymbol{\hat{\tau}_d} \right),
	\end{equation}
\end{linenomath*}
such that the error dynamics are given by
\begin{linenomath*}
	\begin{equation}
		\boldsymbol{\dot{z}} = -\frac{\partial \boldsymbol{\mu}}{\partial \boldsymbol{\nu}}\boldsymbol{M}^{-1}(\boldsymbol{\tau_d}-\boldsymbol{\hat{\tau}_d}) = -\frac{\partial \boldsymbol{\mu}}{\partial \boldsymbol{\nu}}\boldsymbol{M}^{-1}\boldsymbol{z}.  
	\end{equation}
\end{linenomath*}
Generally, the observer dynamics $\boldsymbol{\dot{\zeta}}=\boldsymbol{h}(\boldsymbol{\nu},\boldsymbol{\hat{\tau}_d})$ are formulated as
\begin{linenomath*}
	\begin{equation}
		\boldsymbol{\dot{\zeta}} = -\frac{\partial \boldsymbol{\mu}(\boldsymbol{\nu})}{\partial \boldsymbol{\nu}}\boldsymbol{\dot{\nu}}(\boldsymbol{\tau_d}=\boldsymbol{\zeta}+\boldsymbol{\mu}(\boldsymbol{\nu})),
	\end{equation}
\end{linenomath*}
where the partial derivative $\frac{\partial \boldsymbol{\mu}(\boldsymbol{\nu})}{\partial \boldsymbol{\nu}}$ is given by
\begin{linenomath*}
	\begin{equation}
		\frac{\partial \boldsymbol{\mu}(\boldsymbol{\nu})}{\partial \boldsymbol{\nu}} = \begin{bmatrix}
			\frac{\partial \mu_1}{\partial u} && \frac{\partial \mu_1}{\partial v} && \frac{\partial \mu_1}{\partial r}\\[6pt]
			\frac{\partial \mu_2}{\partial u} && \frac{\partial \mu_2}{\partial v} && \frac{\partial \mu_2}{\partial r}\\[6pt]
			\frac{\partial \mu_3}{\partial u} && \frac{\partial \mu_3}{\partial v} && \frac{\partial \mu_3}{\partial r}
		\end{bmatrix}.
	\end{equation}
\end{linenomath*}
The expression $\boldsymbol{\dot{\nu}}(\boldsymbol{\tau_d}=\boldsymbol{\zeta}+\boldsymbol{\mu}(\boldsymbol{\nu}))$ describes $\boldsymbol{\dot{\nu}}$ as a function of $\boldsymbol{\tau_d}$ defined in \eqref{model_equation} while the equality symbol inside the brackets denotes a substitution.
Considering the error dynamics, and the system dynamics, it is reasonable to define
\begin{linenomath*}
	\begin{equation}
		\boldsymbol{\mu}(\boldsymbol{\nu})=\begin{bmatrix}
			\mu_1(u)\\[6pt]
			\mu_2(v,r)\\[6pt]
			\mu_3(v,r)
		\end{bmatrix},
	\end{equation}
\end{linenomath*}
such that
\begin{linenomath*}
	\begin{equation}
		\frac{\partial \boldsymbol{\mu}(\boldsymbol{\nu})}{\partial \boldsymbol{\nu}} = \begin{bmatrix}
			\frac{\partial \mu_1}{\partial u} && 0 && 0\\[6pt]
			0 && \frac{\partial \mu_2}{\partial v} && \frac{\partial \mu_2}{\partial r}\\[6pt]
			0 && \frac{\partial \mu_3}{\partial v} && \frac{\partial \mu_3}{\partial r}
		\end{bmatrix}.
	\end{equation} 
\end{linenomath*}
With these definitions, the error dynamics yield
\begin{linenomath*}
	\begin{equation}
		\hspace{-0.7em}\resizebox{.9\hsize}{0.04\vsize}{$\boldsymbol{\dot{z}} = -\begin{bmatrix}
				\frac{\partial \mu_1}{\partial u}\kappa_{11} && 0 && 0\\[6pt]
				0 && \frac{\partial \mu_2}{\partial v}\kappa_{22}+\frac{\partial \mu_2}{\partial r}\kappa_{32} && \frac{\partial \mu_2}{\partial v}\kappa_{23}+\frac{\partial \mu_2}{\partial r}\kappa_{33} \\[6pt]
				0 &&  \frac{\partial \mu_3}{\partial v}\kappa_{22}+\frac{\partial \mu_3}{\partial r}\kappa_{32} && \frac{\partial \mu_3}{\partial v}\kappa_{23}+\frac{\partial \mu_3}{\partial r}\kappa_{33}
			\end{bmatrix}\begin{bmatrix}
				z_1\\
				z_2\\
				z_3
			\end{bmatrix}$}.
	\end{equation}
\end{linenomath*}
Regarding $z_1$, it is evident that the error dynamics are exponentially stable for every $\mu_1(u)$, which satisfies
\begin{linenomath*}
	\begin{equation}
		\frac{\partial \mu_1}{\partial u}\kappa_{11} > 0. \label{eq: inequality_gamma1}
	\end{equation}
\end{linenomath*}
To guarantee stable error dynamics concerning $z_2$ and $z_3$, we define the following conditions\footnote{An exclamation mark (!) above an equality (=) or inequality (>) sign means that the expression has to be valid to fulfill a hypothesis.}.
\begin{linenomath*}
	\begin{equation}
		\textbf{C1:}	\ \ \ \ \ \ \ \ \ \ \ \	\frac{\partial \mu_2}{\partial v}\kappa_{23}+\frac{\partial \mu_2}{\partial r}\kappa_{33} \stackrel{!}{=} 0 \label{eq: C1}
	\end{equation} 
	\begin{equation}
		\textbf{C2:}	\ \ \ \ \ \ \ \ \ \ \ \		\frac{\partial \mu_2}{\partial v}\kappa_{22}+\frac{\partial \mu_2}{\partial r}\kappa_{32} \stackrel{!}{>} 0 \label{eq: C2}
	\end{equation} 
	\begin{equation}
		\textbf{C3:}	\ \ \ \ \ \ \ \ \ \ \ \		\frac{\partial \mu_3}{\partial v}\kappa_{22}+\frac{\partial \mu_3}{\partial r}\kappa_{32} \stackrel{!}{=} 0 \label{eq: C3}
	\end{equation} 
	\begin{equation}
		\textbf{C4:}	\ \ \ \ \ \ \ \ \ \ \ \		\frac{\partial \mu_3}{\partial v}\kappa_{23}+\frac{\partial \mu_3}{\partial r}\kappa_{33} \stackrel{!}{>} 0 \label{eq: C4}
	\end{equation} 
\end{linenomath*}
Here, \textbf{C1} and \textbf{C2} guarantee stable error dynamics with regard to $z_2$, while \textbf{C3} and \textbf{C4} ensure stable error dynamics concerning $z_3$. Since \textbf{C1} and \textbf{C3} are stronger conditions, we initially define 
\begin{linenomath*}
	\begin{align}
		\mu_2(v,r) = \Gamma_2\left(\frac{1}{\kappa_{22}}v - \frac{\kappa_{23}}{\kappa_{22}\kappa_{33}}r \right),\\
		\mu_3(v,r) = \Gamma_3\left(\frac{1}{\kappa_{33}}r - \frac{\kappa_{32}}{\kappa_{22}\kappa_{33}}v \right),
	\end{align} 
\end{linenomath*}
where $\Gamma_2$ and $\Gamma_3$ are adaptive gains. With these relations, \textbf{C1} and \textbf{C3} hold. Consequential, \textbf{C2} and \textbf{C4} lead to
\begin{linenomath*}
	\begin{align}
		\Gamma_2\left(1 - \frac{\kappa_{23}\kappa_{32}}{\kappa_{22}\kappa_{33}} \right)\stackrel{!}{>} 0, \label{eq:cond_2}\\
		\Gamma_3\left(1 - \frac{\kappa_{23}\kappa_{32}}{\kappa_{22}\kappa_{33}} \right)\stackrel{!}{>} 0.\label{eq:cond_4}
	\end{align}
\end{linenomath*}
\begin{rmk}
	Since all entries of the mass matrix are positive, $\frac{\kappa_{23}\kappa_{32}}{\kappa_{22}\kappa_{33}}$ is always positive.
\end{rmk}
Concerning \eqref{eq:cond_2} and \eqref{eq:cond_4}, we must consider three cases. 
\\
\\
\textbf{Case 1:} $\kappa_{23}\kappa_{32}<\kappa_{22}\kappa_{33}$
\\
In practice, the entries of the secondary diagonal of the mass matrix are usually much smaller than the diagonal. Hence, the absolute values of $\kappa_{23}$ and $\kappa_{32}$ are also much smaller than $\kappa_{22}$ and $\kappa_{33}$. In this case \textbf{C2} and \textbf{C4} are satisfied if $\Gamma_2>0$ and $\Gamma_3>0$.
\\
\\
\textbf{Case 2:} $\kappa_{23}\kappa_{32}>\kappa_{22}\kappa_{33}$
\\
This case is usually not true. However, if this unrealistic scenario holds, \textbf{C2} and \textbf{C4} are satisfied if $\Gamma_2<0$ and $\Gamma_3<0$.
\\
\\
\textbf{Case 3:} $\kappa_{23}\kappa_{32}=\kappa_{22}\kappa_{33}$
\\
In this case, the observer will not adapt. However, this case is usually neglectable since this case is also unrealistic.

For future considerations, we regard the first case. To satisfy \eqref{eq: inequality_gamma1} and to synchronize the error dynamics, we define
\begin{linenomath*}
	\begin{equation}
		\mu_1(u) = \Gamma_1\frac{1}{\kappa_{11}}u\left(1 - \frac{\kappa_{23}\kappa_{32}}{\kappa_{22}\kappa_{33}} \right). \label{eq: mu_1}
	\end{equation}
\end{linenomath*}
The previous case study also holds for \eqref{eq: mu_1} and the corresponding error dynamics of $z_1$. Therefore, $\Gamma_1>0$ must hold for practical applications.
Summarized, the mapping $\boldsymbol{\mu}(\boldsymbol{\nu})$ is expressed by
\begin{align}
	\boldsymbol{\mu}(\boldsymbol{\nu}) = \begin{bmatrix}
		\mu_1(u)\\[6pt]
		\mu_2(v,r)\\[6pt]
		\mu_3(v,r)
	\end{bmatrix}=\begin{bmatrix}
		\Gamma_1\frac{1}{\kappa_{11}}u\left(1 - \frac{\kappa_{23}\kappa_{32}}{\kappa_{22}\kappa_{33}} \right)\\[6pt]
		\Gamma_2\left(\frac{1}{\kappa_{22}}v - \frac{\kappa_{23}}{\kappa_{22}\kappa_{33}}r \right)\\[6pt]
		\Gamma_3\left(\frac{1}{\kappa_{33}}r - \frac{\kappa_{32}}{\kappa_{22}\kappa_{33}}v\right)
	\end{bmatrix} 
\end{align}

The adaptive gains $\Gamma_1$, $\Gamma_2$, and $\Gamma_3$ define the adaptation speed. These are design parameters and have to be chosen wisely. If $\Gamma_1=\Gamma_2=\Gamma_3$, the estimations of the observed disturbances adapt with the same speed due to the defined error dynamics. Hence, this observer formulation provides a comfortable approach for synchronizing the adaptation speed. With these relations, the error dynamics of the observer are given by
\begin{linenomath*}
	\begin{align}
		\boldsymbol{\dot{z}} = -\begin{bmatrix}
			\Gamma_1\sigma && 0 && 0\\[6pt]
			0 && \Gamma_2\sigma && 0 \\[6pt]
			0 &&  0 && \Gamma_3\sigma
		\end{bmatrix}\begin{bmatrix}
			z_1\\
			z_2\\
			z_3
		\end{bmatrix},
	\end{align}
\end{linenomath*}
where $\sigma=1 - \frac{\kappa_{23}\kappa_{32}}{\kappa_{22}\kappa_{33}}$ leading to globally exponentially stable error dynamics.
Denoting $\boldsymbol{T}= \frac{\partial \boldsymbol{\mu}(\boldsymbol{\nu})}{\partial \boldsymbol{\nu}}$, yields
\begin{linenomath*}
	\begin{equation}
		\boldsymbol{T}=\begin{bmatrix}
			\Gamma_1\frac{1}{\kappa_{11}}\sigma && 0 && 0\\[6pt]
			0 && \Gamma_2\frac{1}{\kappa_{22}} && -\Gamma_2 \frac{\kappa_{23}}{\kappa_{22}\kappa_{33}}\\[6pt]
			0 && -\Gamma_3 \frac{\kappa_{32}}{\kappa_{22}\kappa_{33}} &&\Gamma_3\frac{1}{\kappa_{33}}
		\end{bmatrix}.
	\end{equation}
\end{linenomath*}
Thus, the final expression of the disturbance observer is obtained by
\begin{linenomath*}
	\begin{equation}
		\boldsymbol{\hat{\tau}_d} = \boldsymbol{\zeta}+\boldsymbol{T}\boldsymbol{\nu}, \label{eq:disturbance_observer}
	\end{equation}
\end{linenomath*}
where the observer update is given by
\begin{linenomath*}
	\begin{equation}
		\boldsymbol{\dot{\zeta}} = -\boldsymbol{T}\boldsymbol{\dot{\nu}}(\boldsymbol{\tau_d}=\boldsymbol{\zeta}+\boldsymbol{T}\boldsymbol{\nu}).
	\end{equation}
\end{linenomath*}

Considering the Lyapunov function candidate 
\begin{linenomath*}
	\begin{equation}
		V = \frac{1}{2} \boldsymbol{z}^\top \boldsymbol{z},
	\end{equation}
\end{linenomath*}
it can be shown that even if $\boldsymbol{\dot{\tau}_d} \neq 0$, the error of the disturbance observer converges exponentially into a ball with radius $r_{b} = \frac{\theta}{\sqrt{2\lambda_\mathrm{min}(\boldsymbol{\Gamma})\sigma-1}}$.
\\
\\
\\
\textbf{Proof.}~Defining the adaptive gain matrix $\boldsymbol{\Gamma}=\mathrm{diag}(\Gamma_1, \Gamma_2, \Gamma_3)$, the derivative of the Lyapunov function yields
\begin{linenomath*}
	\begin{equation}
		\dot{V} 
		= - \boldsymbol{z}^\top \boldsymbol{\Gamma}\sigma \boldsymbol{z} + \boldsymbol{z}^\top \boldsymbol{\dot{\tau}_d}. \label{Lyapunov_derivation}
	\end{equation}
\end{linenomath*}
Since
\begin{linenomath*}
	\begin{align}
		0 &\leq (\boldsymbol{z} - \boldsymbol{\dot{\tau}_d})^\top(\boldsymbol{z} - \boldsymbol{\dot{\tau}_d})\notag \\
		&= \boldsymbol{z}^\top\boldsymbol{z} - \boldsymbol{z}^\top\boldsymbol{\dot{\tau}_d} - \boldsymbol{\dot{\tau}_d}^\top\boldsymbol{z} + \boldsymbol{\dot{\tau}_d}^\top \boldsymbol{\dot{\tau}_d}\\
		&=\boldsymbol{z}^\top\boldsymbol{z} - 2\boldsymbol{z}^\top\boldsymbol{\dot{\tau}_d} + \boldsymbol{\dot{\tau}_d}^\top \boldsymbol{\dot{\tau}_d}, \notag
	\end{align}
\end{linenomath*}
\eqref{Lyapunov_derivation} leads to inequality
\begin{linenomath*}
	\begin{align}
		\dot{V} &\leq -\lambda_\mathrm{min}(\boldsymbol{\Gamma})\sigma \boldsymbol{z}^\top \boldsymbol{z} + \frac{1}{2} \boldsymbol{z}^\top \boldsymbol{z} + \frac{1}{2}\boldsymbol{\dot{\tau}_d}^\top \boldsymbol{\dot{\tau}_d}
		\\[0.5pt]
		&\leq -(2\lambda_\mathrm{min}(\boldsymbol{\Gamma})\sigma-1)V + \frac{1}{2}\theta^2, \label{ball_inequation}
	\end{align} 
\end{linenomath*}
where $\lambda_\mathrm{min}(\boldsymbol{\Gamma})$ characterizes the smallest eigenvalue of $\boldsymbol{\Gamma}$ which is likewise the smallest adaptive gain $\mathrm{min}(\Gamma_1,\Gamma_2,\Gamma_3)$, and $\theta$ denotes the maximum possible norm $||\boldsymbol{\dot{\tau}_{d,max}}||$. 
To guarantee stable behavior of \eqref{ball_inequation}, the expression
\begin{linenomath*}
	\begin{equation}
		\lambda_\mathrm{min}(\boldsymbol{\Gamma})\sigma > \frac{1}{2}
	\end{equation}
\end{linenomath*}
has to be satisfied.
The solution of \eqref{ball_inequation} is given by 
\begin{linenomath*}
	\begin{equation}
		0 \leq V(t) \leq \frac{\theta^2}{(4\lambda_\mathrm{min}(\boldsymbol{\Gamma})\sigma-2)}\left(1-e^{-(2\lambda_\mathrm{min}(\boldsymbol{\Gamma})\sigma-1)t} \right).
	\end{equation}
\end{linenomath*}
Hence, $V$ is bounded by $\frac{\theta^2}{(4\lambda_\mathrm{min}(\boldsymbol{\Gamma})\sigma-2)}$ and thus the estimation
error of the disturbance observer converges into a ball 
with radius $r_{b} = \frac{\theta}{\sqrt{2\lambda_\mathrm{min}(\boldsymbol{\Gamma})\sigma-1}}$.

\begin{rmk}
	The stability analysis by using the Lyapunov approach shows likewise to the previous formulations that the error converges exponentially to zero if $\theta=0$, which implies that the disturbances are constant.
\end{rmk}

\subsection{Observer Framework}\label{section: Framework}
\begin{figure}[htb!]
	\centering
	\includegraphics[width=\linewidth]{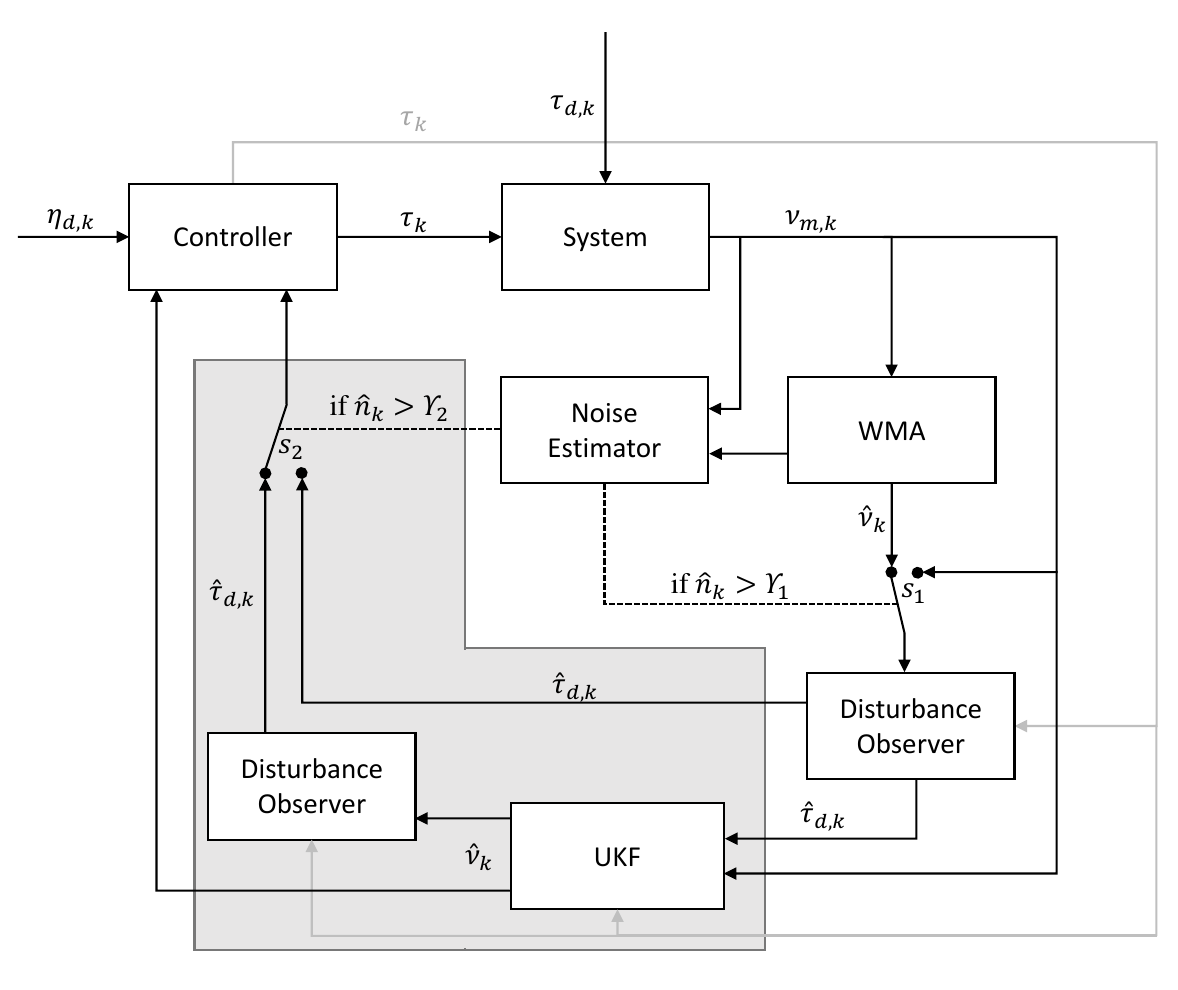}
	\caption{Observer framework}
	\label{fig:Observer_Framework}
\end{figure}

Measurements arrive as a temporal sequence and thus have to be treated as discrete values.
Hence, the discrete measurements at time step $k$ of the system states are defined as
\begin{linenomath*}
	\begin{equation}
		\boldsymbol{\nu}_{m,k}=\boldsymbol{\nu}_k+\boldsymbol{\mathrm{n}}_{k},
	\end{equation}
\end{linenomath*}
which expresses the measurement function \eqref{UKF_measurement} with Gaussian noise \eqref{meas_noise}.
The discretized observer is obtained by
\begin{linenomath*}
	\begin{equation}
		\boldsymbol{\hat{\tau}_{d,k}} = \boldsymbol{\zeta}_k+\boldsymbol{T}\boldsymbol{\hat{\nu}}_k,
	\end{equation}
\end{linenomath*}
where $\boldsymbol{\hat{\nu}}_k$ describe the filtered measurements, and the observer update is given by
\begin{linenomath*}
	\begin{equation}
		\boldsymbol{\zeta}_{k+1} = \boldsymbol{\zeta}_k + \Delta t\left( -\boldsymbol{T}\boldsymbol{\nu}_{k+1}(\boldsymbol{\tau_{d,k}}=\boldsymbol{\zeta}_k+\boldsymbol{T}\boldsymbol{\hat{\nu}}_k)\right).
	\end{equation}
\end{linenomath*}
Furthermore, the discrete model of \eqref{model_equation} yields
\begin{linenomath*}
	\begin{equation}
		\hspace{-0.7em}\resizebox{.9\hsize}{0.0145\vsize}{$\boldsymbol{\nu}_{k+1} = \boldsymbol{\nu}_{k} + \Delta t\left( \boldsymbol{M}^{-1}\left(\boldsymbol{\tau}_{k}+\boldsymbol{\tau_{d,k}}-\boldsymbol{D}(\boldsymbol{\nu}_{k})\boldsymbol{\nu}_{k}-\boldsymbol{C}(\boldsymbol{\nu}_{k})\boldsymbol{\nu}_{k}\right)\right)+\boldsymbol{\mathrm{w}}_{k}$},
	\end{equation}
\end{linenomath*}
where $\Delta t$ is the time step between each measurement, and $\boldsymbol{\mathrm{w}}_{k}$ describes the model uncertainty defined in expression \eqref{model_noise}.
However, considering realistic hydrodynamics of vessels, adding Gaussian noise additive to the model dynamics leads to an impulsive unrealistic behavior. Therefore, we assume that the identified hydrodynamic parameters and mass parameters of \tabref{tab: Parameters} are erroneous. For this purpose, the actual model dynamics are simulated with randomly generated parameters according to
\begin{equation}
	p = \hat{p}(1+\rho\mathrm{w}), \label{model_uncertainties}
\end{equation}
where $p$ describes the real parameters, $\hat{p}$ are the identified parameters described in Section \ref{sec:methodandsetup}, $\rho$ is a scaling parameter defining the uncertainty, and $\mathrm{w}$ is a randomly generated Gaussian distributed number. As a result, the model uncertainties affect the model in each model matrix.
The robustness of the disturbance observer concerning measurement noise depends on the current states. If the velocities are very low, high noise has a much stronger effect on the observed states. To identify the relative strength of the noise, we introduce an online noise estimator. It is expressed by
\begin{equation}
	\hspace{-0.75em}\resizebox{.9\hsize}{0.028\vsize}{$\hat{n}_k = \boldsymbol{E}\left[\frac{\left|\left|\sqrt{\boldsymbol{E}\left[\left((\boldsymbol{\nu}_{WMA,k-w:k} - \boldsymbol{\nu}_{m,k-w:k})-\boldsymbol{E}\left[\boldsymbol{\nu}_{WMA,k-w:k} - \boldsymbol{\nu}_{m,k-w:k}\right]\right)^2\right]}\right|\right|}{\left|\left|\boldsymbol{E}\left[\boldsymbol{\nu}_{m,k-w:k}\right]\right|\right|}\right]_{k-4w:k}$}, \label{noise_estimator}
\end{equation}
where $\boldsymbol{E}[\cdot]$ describes the expected value, and $w$ is the considered window size. The foundation of the estimator builds the difference between the actual measurements and their smoothed version provided by its WMA. Hence, the nominator is expressed by the $l^2$-norm of the standard deviation concerning the difference between the measurements and their WMA-filtered version. At the same time, the denominator characterizes the $l^2$-norm of the measurement's mean. As a result, the estimator represents the noise influence depending on the current states. The outer expectation bracket has the purpose of smoothing the estimations. Otherwise, they can be noisy, as depicted in the results.
The entire framework consists of multiple components and has a cascading structure. First, the measurements are propagated through the noise estimator defined in \eqref{noise_estimator}. If the detected measurement noise is neglectable small, the observer framework provides estimations of the disturbances without filtering. Otherwise, the noise estimator triggers a cascaded filtering mechanism depicted in \figref{fig:Observer_Framework}. 

Here, $\gamma_1$ and $\gamma_2$ are the thresholds for triggering the switches $s_1$ and $s_2$. If $\hat{n} > \gamma_1$, the disturbance observer receives the WMA filtered states. In any switch setting of $s_1$, the UKF estimates the actual states by receiving the measurements $\nu_{m,k}$ and the estimated disturbances $\hat{\tau}_{d,k}$. If $\hat{n} > \gamma_2$, a second observer generates smoother versions of $\hat{\tau}_{d,k}$ by receiving the filtered states from the UKF. Note, that reducing the adaptive gains $\Gamma_i$ results in a smoothed estimation of the disturbances. For that reason, the adaptive gains for the second observer should be chosen small. The grey unit in \figref{fig:Observer_Framework} can be duplicated and added to the cascaded structure if the measurement noise is exaggerated.

For traceability, in the following, pseudocode is presented. The code should give clear instructions to reconstruct the entire framework. 
\begin{algorithm}[h]
	\caption{Pseudocode}
	\begin{algorithmic}[1]
		\While{$\mathrm{observation} = \mathrm{True}$}
		\State Calculate $\boldsymbol{\nu}_{WMA,k}$ according to \eqref{eq:WMA}
		\smallskip
		\State Calculate $\hat{n}_{k}$ according to \eqref{noise_estimator}
		\smallskip
		\If{$\hat{n}_{k}\leq\gamma_1$}
		\smallskip
		\State $\boldsymbol{T}(\Gamma_i) = \boldsymbol{T}(\Gamma_{i,1})$ 
		\smallskip
		\State	\hspace{-0.5em}$\boldsymbol{\hat{\tau}_{d,k}} = \boldsymbol{\zeta}_k+\boldsymbol{T}\boldsymbol{\nu}_{m,k}$
		\smallskip
		\State \hspace{-0.5em}$\boldsymbol{\zeta}_{k+1} = \boldsymbol{\zeta}_k + \Delta t\left( -\boldsymbol{T}\boldsymbol{\nu}_{k+1}(\boldsymbol{\tau_{d,k}}=\boldsymbol{\zeta}_k+\boldsymbol{T}\boldsymbol{\nu}_{m,k})\right)$
		\smallskip
		\ElsIf{$\hat{n}_{k}>\gamma_1$}
		\smallskip
		\State $\boldsymbol{T}(\Gamma_i) = \boldsymbol{T}(\Gamma_{i,1})$ 
		\smallskip
		\State	\hspace{-0.5em}$\boldsymbol{\hat{\tau}_{d,k}} = \boldsymbol{\zeta}_k+\boldsymbol{T}\boldsymbol{\nu}_{WMA,k}$
		\smallskip
		\State \hspace{-0.5em}$\boldsymbol{\zeta}_{k+1} = \boldsymbol{\zeta}_k + \Delta t\left( -\boldsymbol{T}\boldsymbol{\nu}_{k+1}(\boldsymbol{\tau_{d,k}}=\boldsymbol{\zeta}_k+\boldsymbol{T}\boldsymbol{\nu}_{WMA,k})\right)$
		\smallskip		
		\ElsIf{$\hat{n}_{k}>\gamma_2$}
		\smallskip
		\State Calculate $\boldsymbol{\mathcal{X}}_{i,k}$ according to \eqref{sigma_points1}-\eqref{sigma_points3}
		\smallskip
		\State Calculate $\boldsymbol{\hat{\nu}}_{a,k+1}$ according to \eqref{UKF_start}-\eqref{UKF_end}
		\smallskip
		\State $\boldsymbol{T}(\Gamma_i) = \boldsymbol{T}(\Gamma_{i,2})$  
		\smallskip
		\State	\hspace{-0.5em}$\boldsymbol{\hat{\tau}_{d,k}} = \boldsymbol{\zeta}_k+\boldsymbol{T}\boldsymbol{\hat{\nu}}_{a,k+1}$
		\smallskip
		\State \hspace{-0.5em}$\boldsymbol{\zeta}_{k+1} = \boldsymbol{\zeta}_k + \Delta t\left( -\boldsymbol{T}\boldsymbol{\nu}_{k+1}(\boldsymbol{\tau_{d,k}}=\boldsymbol{\zeta}_k+\boldsymbol{T}\boldsymbol{\hat{\nu}}_{a,k+1})\right)$
		\smallskip		
		\ElsIf{$\hat{n}_{k}>\gamma_3$}
		\smallskip
		\State Calculate $\boldsymbol{\mathcal{X}}_{i,k}$ according to \eqref{sigma_points1}-\eqref{sigma_points3}
		\smallskip
		\State Calculate $\boldsymbol{\hat{\nu}}_{b,k+1}$ according to \eqref{UKF_start}-\eqref{UKF_end}
		\smallskip
		\State $\boldsymbol{T}(\Gamma_i) = \boldsymbol{T}(\Gamma_{i,3})$ 
		\smallskip
		\State	\hspace{-0.5em}$\boldsymbol{\hat{\tau}_{d,k}} = \boldsymbol{\zeta}_k+\boldsymbol{T}\boldsymbol{\hat{\nu}}_{b,k+1}$
		\smallskip
		\State \hspace{-0.5em}$\boldsymbol{\zeta}_{k+1} = \boldsymbol{\zeta}_k + \Delta t\left( -\boldsymbol{T}\boldsymbol{\nu}_{k+1}(\boldsymbol{\tau_{d,k}}=\boldsymbol{\zeta}_k+\boldsymbol{T}\boldsymbol{\hat{\nu}}_{b,k+1})\right)$
		\smallskip
		\EndIf		
		\EndWhile
	\end{algorithmic}
\end{algorithm}
Initialization parameters of the observer framework are depicted in \tabref{tab:Initialization}. Tuning parameters such as $\Gamma_1$, $\Gamma_2$, $\Gamma_3$, $\alpha$, $\beta$, and $\kappa$, used for this study, are just a recommendation. Here, $\Gamma_{i,j}$ describes the adaptive gains at the $j$-th observer depending on the number of cascading units. In the provided algorithm, two grey units are cascaded, where the state estimations of the first and second UKF are denoted as $\boldsymbol{\hat{\nu}}_{a,k+1}$ and $\boldsymbol{\hat{\nu}}_{b,k+1}$.
Note that the adaptive gains should be adjusted if the sensor measurements have a low sampling frequency since higher sampling rates allow greater adaptive gains. Otherwise, the observations can turn unstable, described in more detail in the following section. 

\section{Method and setup} \label{sec:methodandsetup}
This study is conducted with the specifications of the milliAmpere ferry.
MilliAmpere is Norway's first driverless ferry and part of the Autoferry project of the Norwegian University of Science and Technology \cite{Brekke2022}. 

\subsection{Ferry specification} 
The hydrodynamic parameters and the entries of the mass matrix are identified in \cite{Pedersen2019} by using datasets consisting of velocity measurements $\boldsymbol{\nu}_m$ of the vessel, rotational velocity measurements of the propellers, and azimuth angle measurements of the thrusters. For this purpose, techniques such as optimal control, regularization, and cross-validation are used. The identified parameters are shown in \tabref{tab: Parameters}. 

\subsection{Scenario description}\label{sec: Scenario_description}
The observer framework was tested in several scenarios corresponding to different severity levels of the environmental disturbances.
To evaluate the capability of the framework under highly dynamic conditions, the forces induced by wind, waves, and sea currents are simulated by
\begin{linenomath*}
	\begin{subequations}
		\begin{align}
			&\boldsymbol{\tau}_{\mathrm{wind}} = \begin{bmatrix}
				\bar{F}_{\mathrm{wind}}\cos(\gamma_{\mathrm{wind}}-\psi)\left(1+\sin(t)\right)\\[3pt]
				-\bar{F}_{\mathrm{wind}}\sin(\gamma_{\mathrm{wind}}-\psi)\left(1+\sin(t)\right)\\[3pt]
				\bar{F}_{\mathrm{wind}}\sin(2(\gamma_{\mathrm{wind}}-\psi))\left(1+\sin(t)\right)\frac{L}{4}
			\end{bmatrix}, \label{tau_wind}\\
			&\boldsymbol{\tau}_{\mathrm{wave}} = \begin{bmatrix}
				\bar{F}_{\mathrm{wave}}\cos(\gamma_{\mathrm{wave}}-\psi)\left(1+\sin(t)\right)\\[3pt]
				-\bar{F}_{\mathrm{wave}}\sin(\gamma_{\mathrm{wave}}-\psi)\left(1+\sin(t)\right)\\[3pt]
				0
			\end{bmatrix},\\
			&\boldsymbol{\tau}_{\mathrm{current}} = \begin{bmatrix}
				\bar{F}_{\mathrm{current}}\cos(\gamma_{\mathrm{current}}-\psi)\left(1-e^{-\frac{t}{T_s}})\right)\\[3pt]
				-\bar{F}_{\mathrm{current}}\sin(\gamma_{\mathrm{current}}-\psi)\left(1-e^{-\frac{t}{T_s}})\right)\\[3pt]
				0 
			\end{bmatrix}.\label{tau_current}
		\end{align}
	\end{subequations}
\end{linenomath*}
Note that the impact angles $\gamma_i$ of wind, waves, and currents are related to the $x$-axis of the global coordinate system, while $\bar{F}_i$ are the dedicated mean forces. Moreover, $L$ is the length of the vessel and $T_s$ expresses a time constant. Since the waves are affected by wind, they have approximately the same direction. Furthermore, it is expected that the torque induced by the environment is mainly influenced by the wind. Hence, it is assumed that the wind force in the sway direction describes the torque dependent on $\cos(\gamma_{\mathrm{wind}}-\psi)$ impacting half of the vessel's length. This inference follows from geometrical relations and the assumption that the vessel's hull is approximately symmetrical from the center to the stern side and from the center to the nose side. Note that the identity
\begin{equation}
	\cos(\gamma_{\mathrm{wind}}-\psi)\sin(\gamma_{\mathrm{wind}}-\psi)=\frac{1}{2}\sin(2(\gamma_{\mathrm{wind}}-\psi))
\end{equation}
is valid. In addition, an oscillating force is superposed to the wind's mean force in order to simulate a pulsating wind behavior. While the waves are simulated as an oscillating force, it is assumed that the sea currents do not exhibit highly dynamic behavior. Therefore, they are simulated as an exponentially decaying force.

The entire environmental influence described by \eqref{tau_wind}-\eqref{tau_current} results in highly dynamic disturbances affecting the vessel. In addition, the scenario is made more challenging by adding strongly increasing measurement noise.
To visualize the environmental impact, the control input $\tau$ is set to zero. The simulation lasts for $\unit[100]{s}$, and the measurement noise is increased stepwise. This has the purpose of showcasing the capability of the framework under different conditions. Hence, the covariance matrix of the measurement noise is defined as
\begin{align}
	\boldsymbol{R} = \begin{cases}
		\boldsymbol{0},\ \ \ \ \ \ \mathrm{if}\ \ \ \unit[0]{s} \leq t < \unit[20]{s}, \\
		\boldsymbol{R}_1,\ \ \ \ \mathrm{if}\ \ \ \unit[20]{s} \leq t < \unit[40]{s}, \\
		\boldsymbol{R}_2,\ \ \ \ \mathrm{if}\ \ \ \unit[40]{s} \leq t < \unit[60]{s}, \\
		\boldsymbol{R}_3,\ \ \ \ \mathrm{if}\ \ \ \unit[60]{s} \leq t < \unit[80]{s}, \\
		\boldsymbol{R}_4,\ \ \ \ \mathrm{if}\ \ \ \unit[80]{s} \leq t \leq \unit[100]{s}, \\
	\end{cases}
\end{align}
The simulation parameters are depicted in \tabref{tab: Simulation parameters}. 
Under consideration of measurement uncertainties, it turned out that the observer mainly issues with scenarios where the measured states have small values since they are more sensitive to noise. If the environmental disturbances are severe, leading to a higher force impacting the vessel, measurement noise has a minor influence on the system states since the forces have greater absolute values and are thus easier to handle by the observer. 
Considering \eqref{eq:disturbance_observer}, this statement is reasonable since the value of the observer variable $\boldsymbol{\zeta}$ will dominate the influence of the measurement noise within $\boldsymbol{\nu}$ if the disturbances are severe. Various simulations verified this assumption. 

In addition, simulations have shown that the adaptive gains $\Gamma_i$ should be adjusted to the measurement time step $\Delta t$. A higher sampling frequency, i.e., a lower time step, offers the possibility to increase the adaptive gains very high. However, reducing the sampling frequency results in a reciprocal effect. If the adaptive gains are improperly adjusted to the sampling frequency, the observer turns unstable. Despite that, a time step of $\Delta t=\unit[0.1]{s}$ still enables $\Gamma_i=18$, which is absolutely sufficient for a fast adaptation. While higher adaptive gains lead to faster adaptation, decreasing the adaptive gains is beneficial for generating smoothed estimations. Both of these advantages are used in the framework showcased in \figref{fig:Observer_Framework}.
The parameterization of the observer framework is given in \tabref{tab:Initialization}. To adapt fast enough to the real values, the first observer has higher adaptive gains. The subsequent cascaded observers have lower gains since their main function is to smooth noisy signals. Hence, the parameterization of switch $s_3$ is chosen such that the noise rejection dominates the adaptation speed and shall demonstrate an emergency trigger.
Tuning parameters $\gamma_i$ of the switches were chosen according to the best generalizability concerning multiple simulated scenarios, including several severity levels of the disturbances and measurement noise.
Furthermore, the window size $w$ should be adapted to the sampling frequency. Simulations showed that a proper calibration is done by choosing $w=\frac{1}{\Delta t}$.

To compare the influence of various $\Gamma_i$ and model uncertainties, a second simulation is showcased according to
\begin{equation}
	\boldsymbol{\tau}_{d} = \begin{bmatrix}
		||\boldsymbol{\tau}_{\mathrm{current}}||\\
		||\boldsymbol{\tau}_{\mathrm{current}}||\\
		||\boldsymbol{\tau}_{\mathrm{current}}||
	\end{bmatrix}. \label{sim_2}
\end{equation}
Furthermore, the second simulation is conducted with a very low measurement sampling rate of $\Delta t = \unit[0.1]{s}$ to showcase the observer's capability of dealing with low sampling frequencies. To compare the behavior of the observer framework under uncertain model circumstances, model uncertainties are simulated according to \eqref{model_uncertainties} with various $\rho$ given in \tabref{tab: Simulation parameters}. Note that $\rho=0.1$ corresponds to a model parameter uncertainty of 10\%.

\section{Results and discussions}\label{section: Results}
\begin{figure*}
	\centering
	\begin{subfigure}{0.46\linewidth}
		\includegraphics[width=\linewidth]{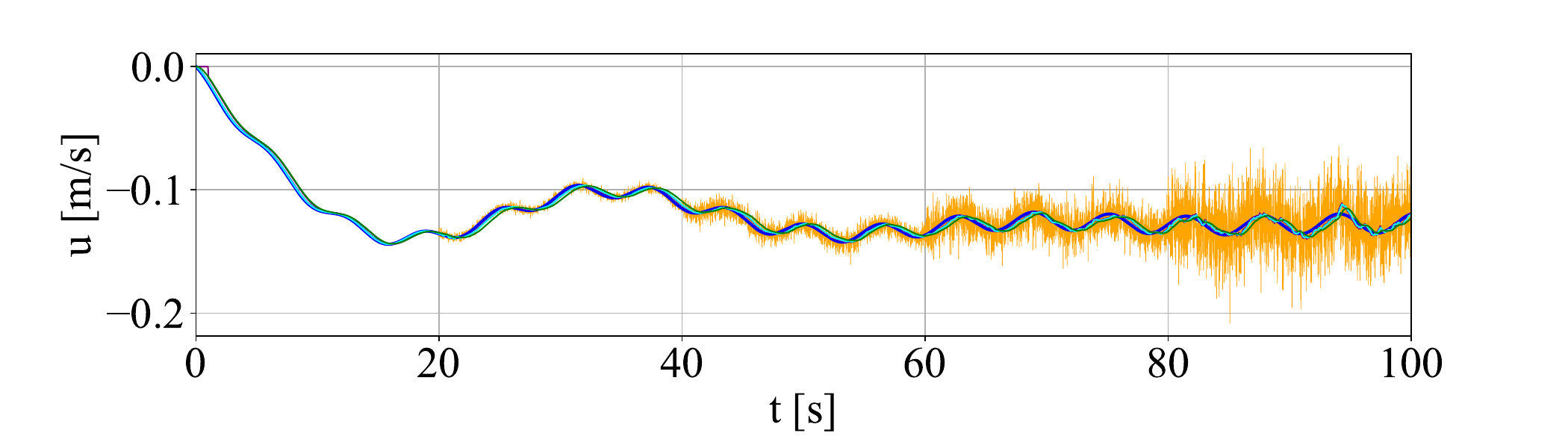}
		\caption{Surge}
	\end{subfigure}
	\begin{subfigure}{0.46\linewidth}
		\includegraphics[width=\linewidth]{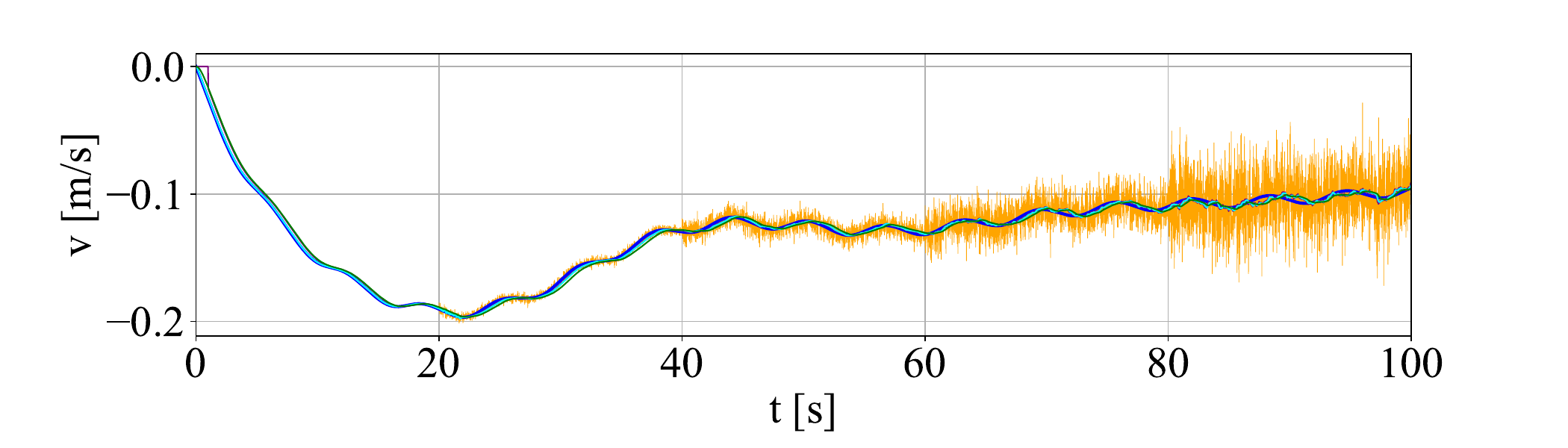}
		\caption{Sway}
	\end{subfigure}\\
	\begin{subfigure}{0.46\linewidth}
		\includegraphics[width=\linewidth]{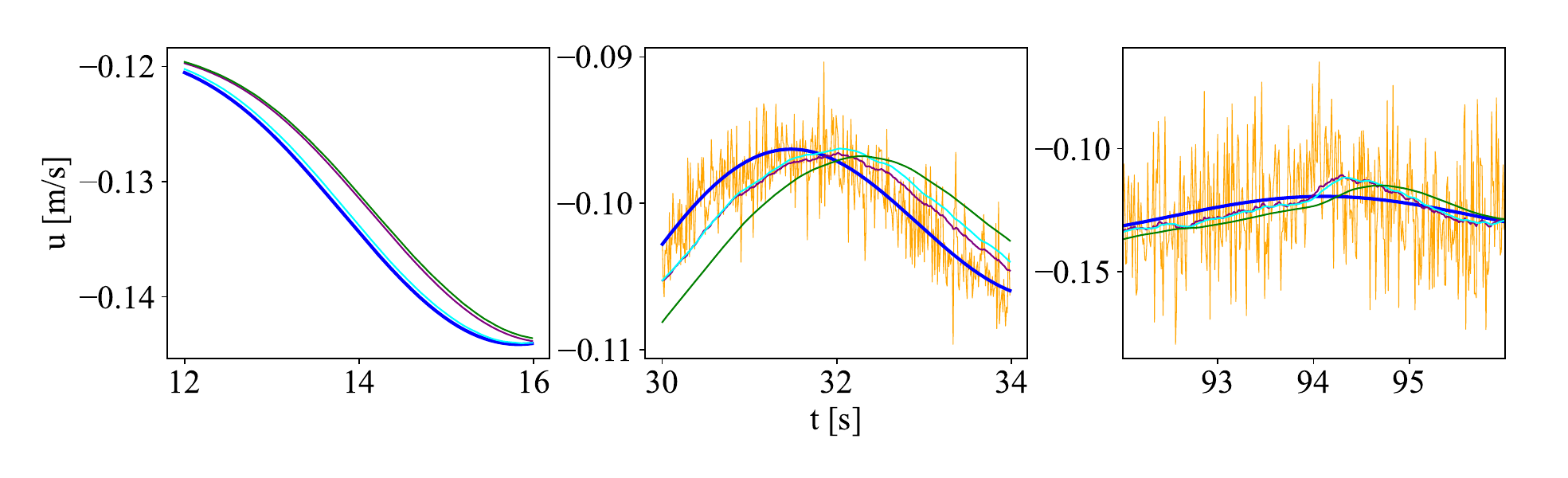}
		\caption{Fractions of surge}
	\end{subfigure}
	\begin{subfigure}{0.46\linewidth}
		\includegraphics[width=\linewidth]{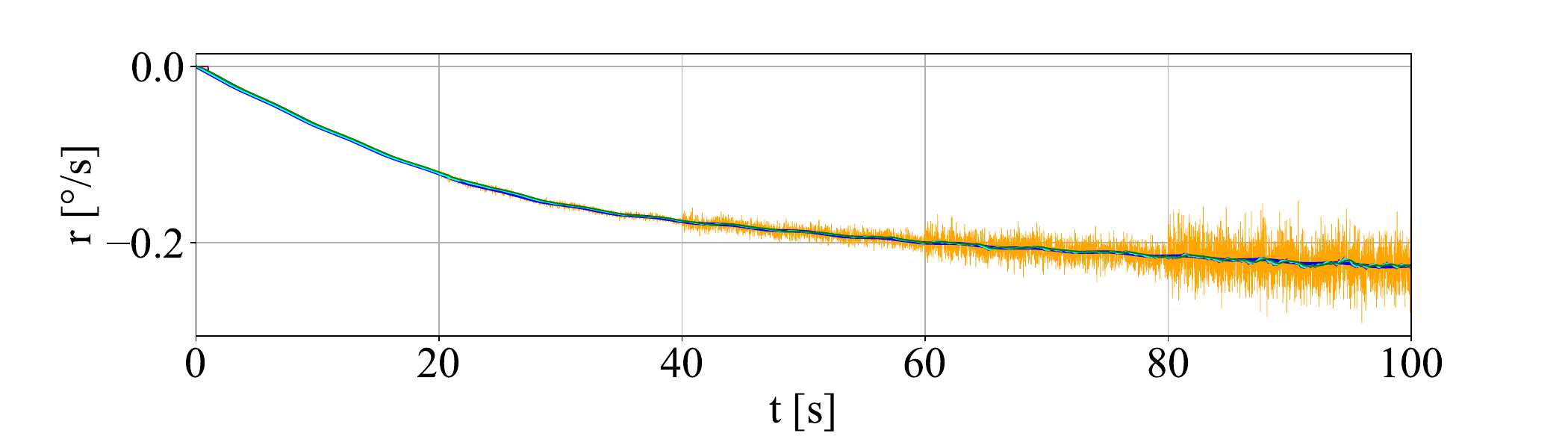}
		\caption{Rotational speed}
	\end{subfigure}
	\caption{Real states (\protect\blueline), measured states with measurement noise (\protect\orangeline), WMA filtered measurements (\protect\purpleline), filtered states from the first UKF (\protect\cyanline), and filtered states from the second UKF (\protect\greenline).}
	\label{fig: States}
\end{figure*}

In \figref{fig: States}, the real system states, the measured states, the WMA filtered measurements, and the filtered measurements from the UKFs are visualized. While the WMA filtering seems to provide sufficient smoothing for weak measurement noise, smooth state estimations under consideration of intense measurement noise can only be reasonably generated by the following cascaded UKFs. 

\begin{figure*}
	\centering
	\begin{subfigure}{0.46\linewidth}
		\includegraphics[width=\linewidth]{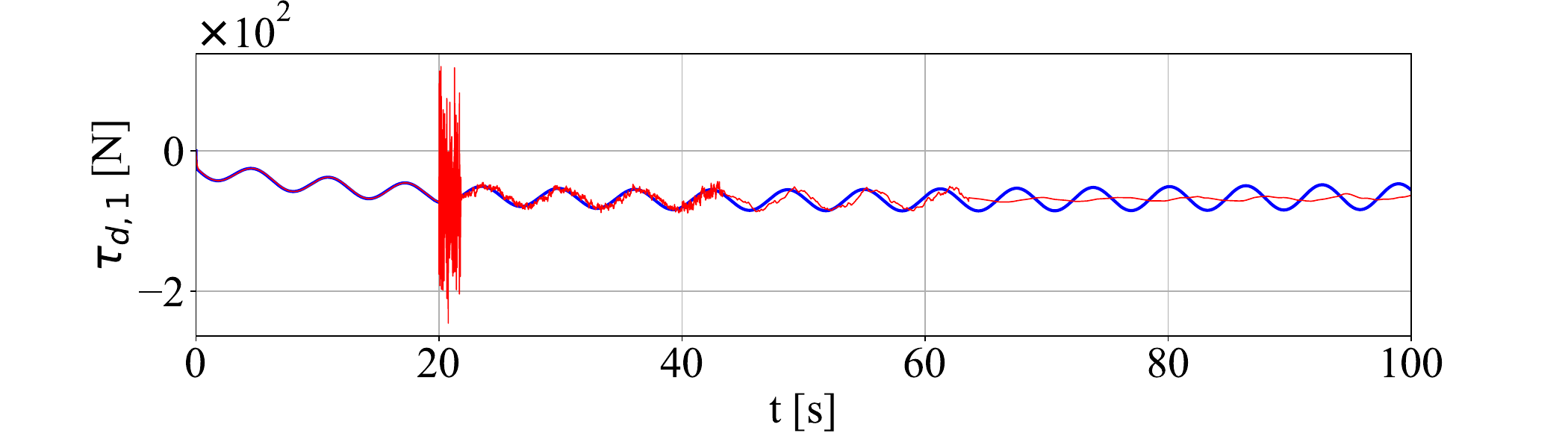}
		\caption{Observed disturbances concerning surge}
	\end{subfigure}
	\begin{subfigure}{0.46\linewidth}
		\includegraphics[width=\linewidth]{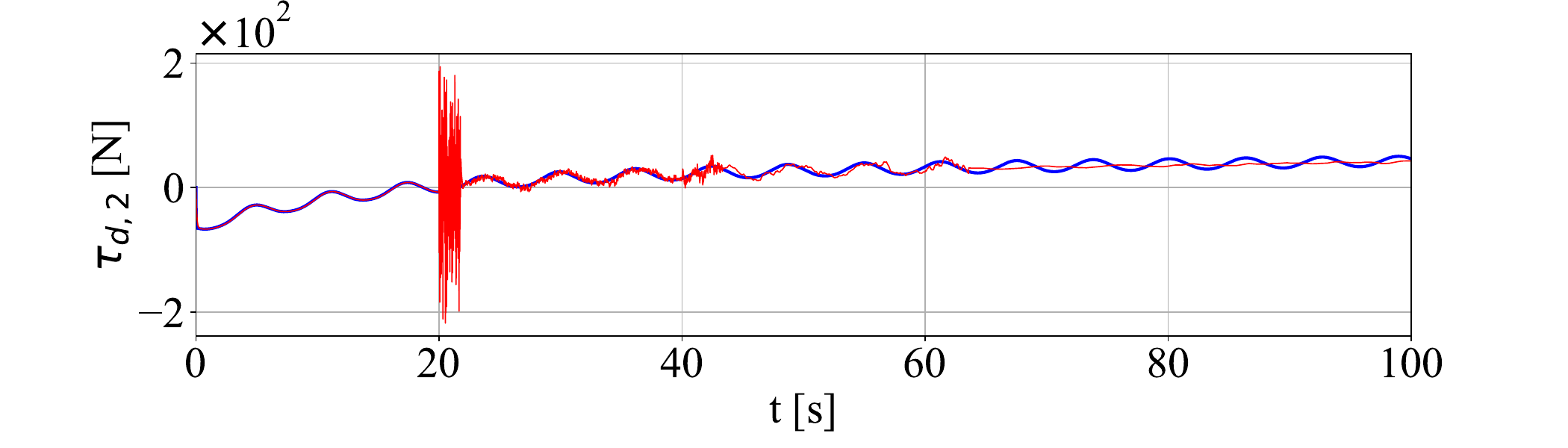}
		\caption{Observed disturbances concerning sway}
	\end{subfigure}\\
	\begin{subfigure}{0.46\linewidth}
		\includegraphics[width=\linewidth]{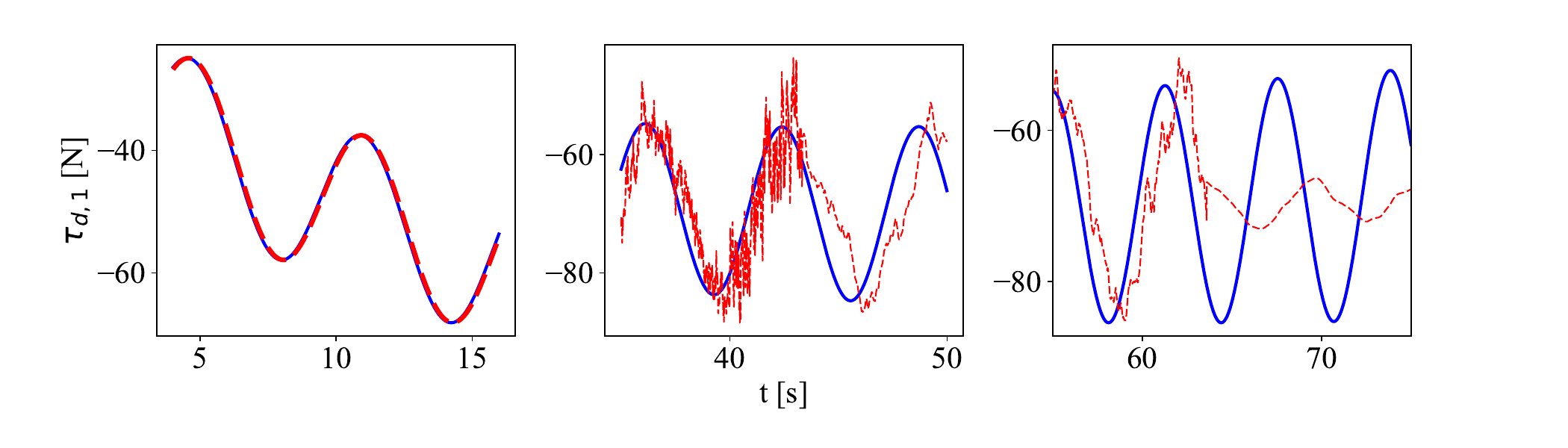}
		\caption{Fractions of the observed disturbance $\tau_{d,1}$}
	\end{subfigure}
	\begin{subfigure}{0.46\linewidth}
		\includegraphics[width=\linewidth]{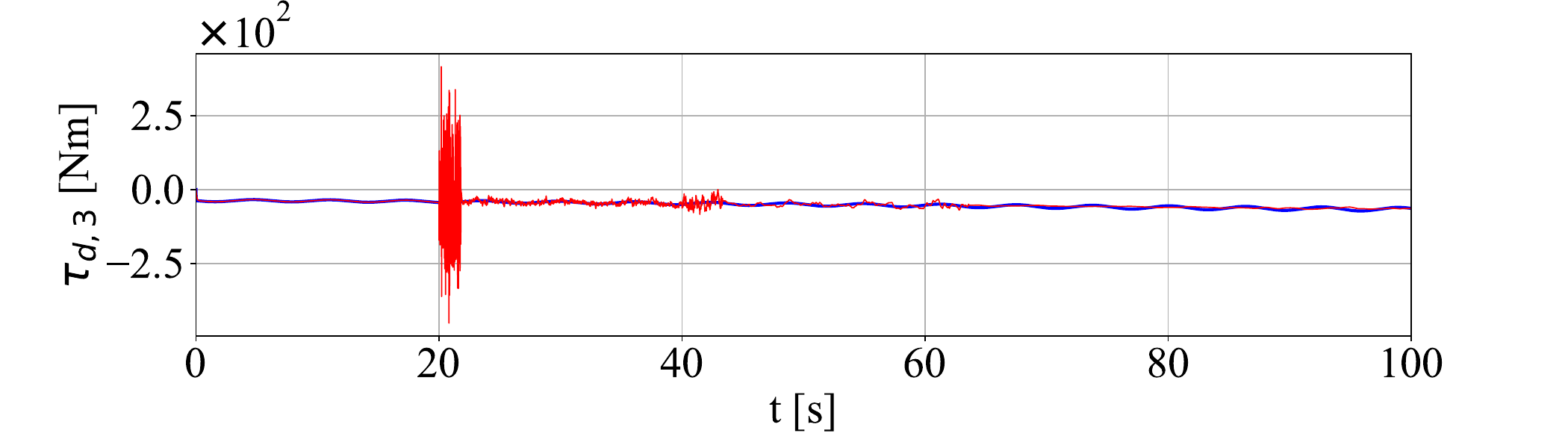}
		\caption{Observed disturbances concerning rotational speed}
	\end{subfigure}
	\caption{Observed disturbances $\boldsymbol{\hat{\tau}_d}$ (\protect\redline) compared to the real disturbances $\boldsymbol{\tau_d}$ (\protect\blueline).}
	\label{fig: Observer}
\end{figure*}

\figref{fig: Observer} depicts the estimated disturbances from the observer framework. The first twenty seconds are simulated without measurement noise. Hence, the observer is adapting in real-time highly accurate. However, one can see at $\unit[20]{s}<t<\unit[22]{s}$ how even weak measurement noise propagates through the observer. The WMA smoothing of the observed disturbances is a bit delayed since the noise estimator needs a short time to adapt. 

\begin{figure*}
	\centering
	\begin{subfigure}{0.46\linewidth}
		\includegraphics[width=\linewidth]{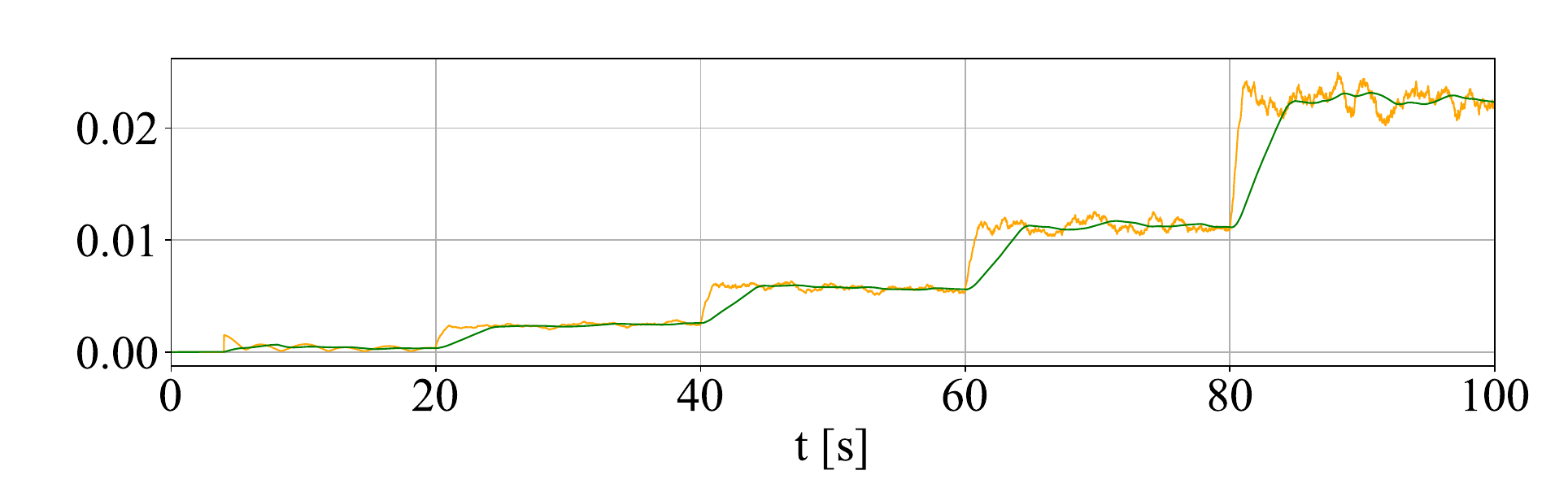}
		\caption{Noise estimation}
	\end{subfigure}
	\begin{subfigure}{0.46\linewidth}
		\includegraphics[width=\linewidth]{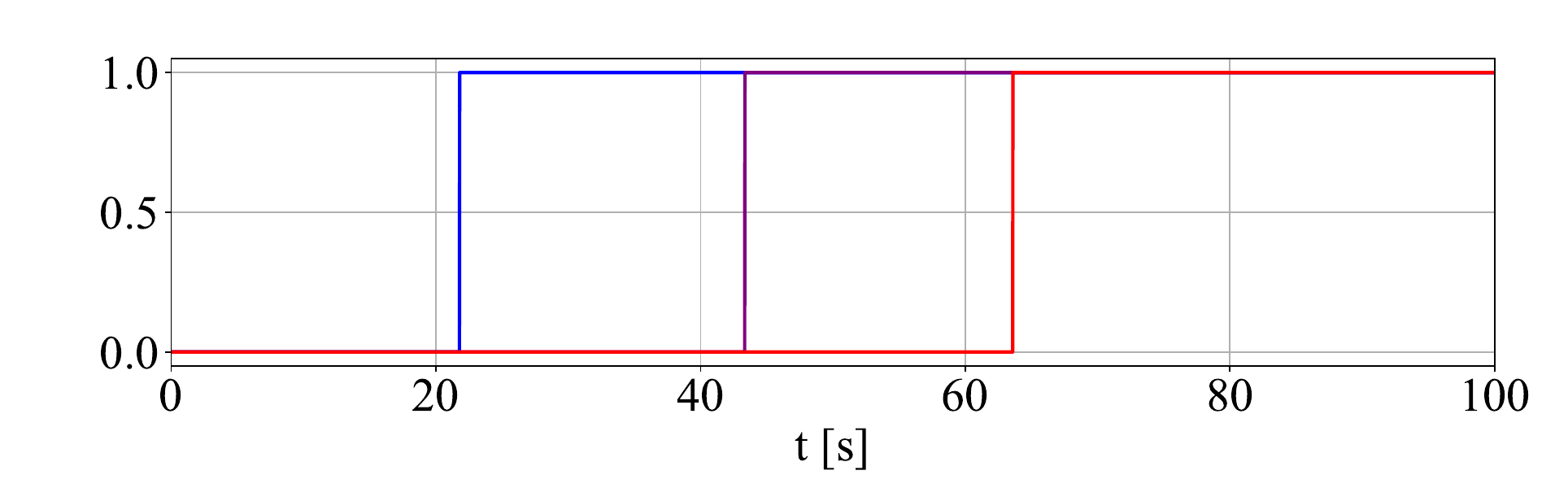} 
		\caption{Switches}
	\end{subfigure}
	\caption{Noise estimation without mean smoothing (\protect\orangeline), and the final smoothed noise strength estimation (\protect\greenline) triggering the switches $s_1$ (\protect\blueline), $s_2$ (\protect\purpleline), and $s_3$ (\protect\redline).}
	\label{fig:Noise_estimator_and_switches}
\end{figure*}

In \figref{fig:Noise_estimator_and_switches}, the noise estimations and the triggered switches are visualized. Here, $s_1$ and $s_2$ are according to the representation in \figref{fig:Observer_Framework}, while the third switch $s_3$ is an additional cascading unit depicted in grey. If the measurement noise is getting too intense, an additional smoothing unit is connected. Usually, measurements showcased at $t>\unit[60]{s}$ are considered as unrealistically uncertain. However, the framework allows a smoothed estimation even if the disturbances are highly dynamic. If the disturbances are less dynamic, the smoothed estimations will adapt in real-time to the actual values. 

\begin{figure*}
	\centering
	\begin{subfigure}{0.46\linewidth}
		\includegraphics[width=\linewidth]{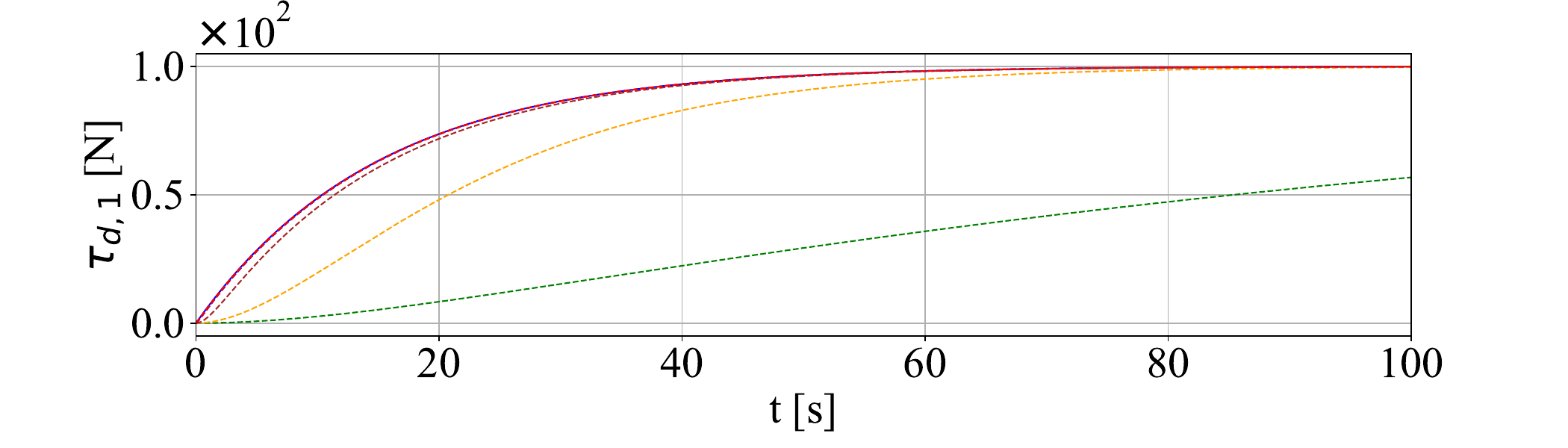} 
		\caption{Observed disturbances concerning surge}
	\end{subfigure}
	\begin{subfigure}{0.46\linewidth}
		\includegraphics[width=\linewidth]{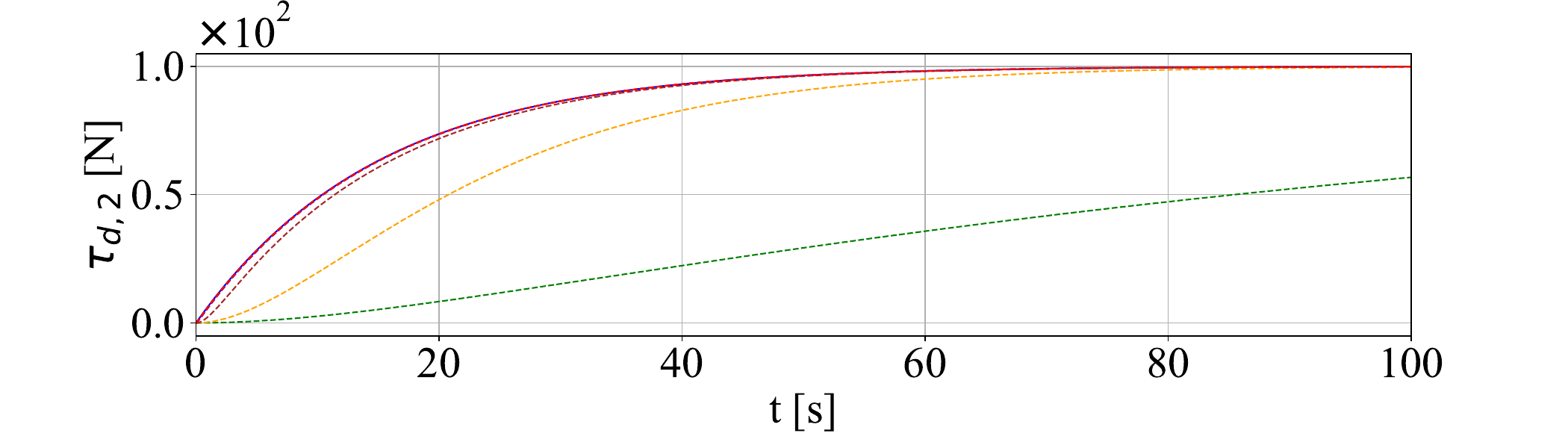}
		\caption{Observed disturbances concerning sway}
	\end{subfigure}\\
	\begin{subfigure}{0.46\linewidth}
		\includegraphics[width=\linewidth]{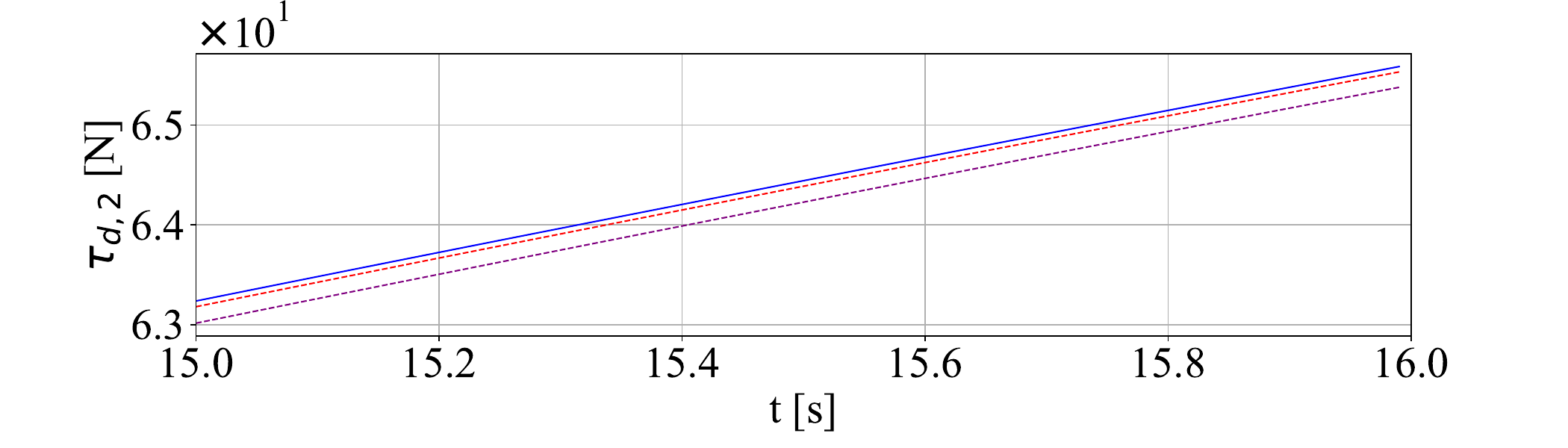} 
		\caption{Fractions of the observed disturbance $\tau_{d,1}$}
	\end{subfigure}
	\begin{subfigure}{0.46\linewidth}
		\includegraphics[width=\linewidth]{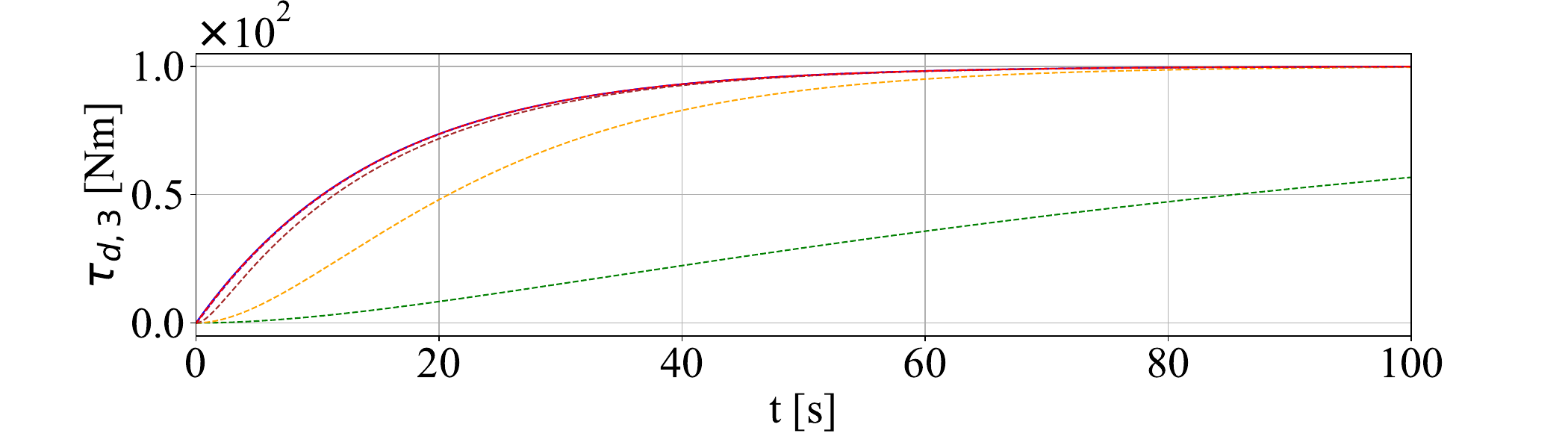} 
		\caption{Observed disturbances concerning rotational speed}
	\end{subfigure}
	\caption{Comparison of different $\Gamma_i$, for $i=1,2,3$ with $\Gamma_i = 0.01$ (\protect\greenlinedashed), $\Gamma_i = 0.1$ (\protect\orangelinedashed), $\Gamma_i = 1$ (\protect\brownlinedashed), $\Gamma_i = 10$ (\protect\purplelinedashed), $\Gamma_i = 18$ (\protect\redlinedashed), and the real disturbances (\protect\blueline). This simulation is designed such that all disturbances represent equal behavior for the purpose of better comparability. Here, it is shown that the designed observer allows a convenient adjustment of the adaptive gains.}
	\label{fig: Gamma comparison}
\end{figure*}

\figref{fig: Gamma comparison} compares the various adaptation gains $\Gamma_i$ conducted with the simulation described by \eqref{sim_2}. One can see that an increase of $\Gamma_i$ leads to faster adaptations, while lower $\Gamma_i$ smoothens the estimations. Furthermore, the synchronization of the observer is evident.
The observer framework benefits from synchronizing the observed disturbances since each cascaded unit ensures an equal adaptation speed and smoothing property.

\begin{figure*}
	\centering
	\begin{subfigure}{0.33\linewidth}
		\includegraphics[width=\linewidth]{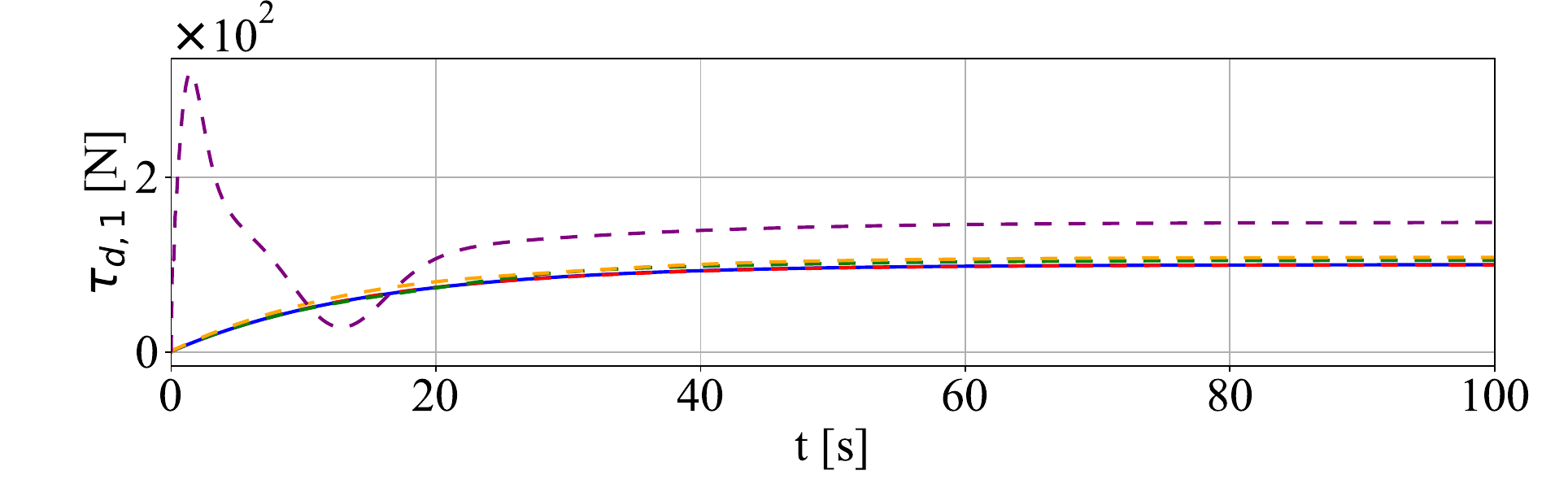} 
		\caption{Observed disturbances concerning surge}
	\end{subfigure}
	\begin{subfigure}{0.33\linewidth}
		\includegraphics[width=\linewidth]{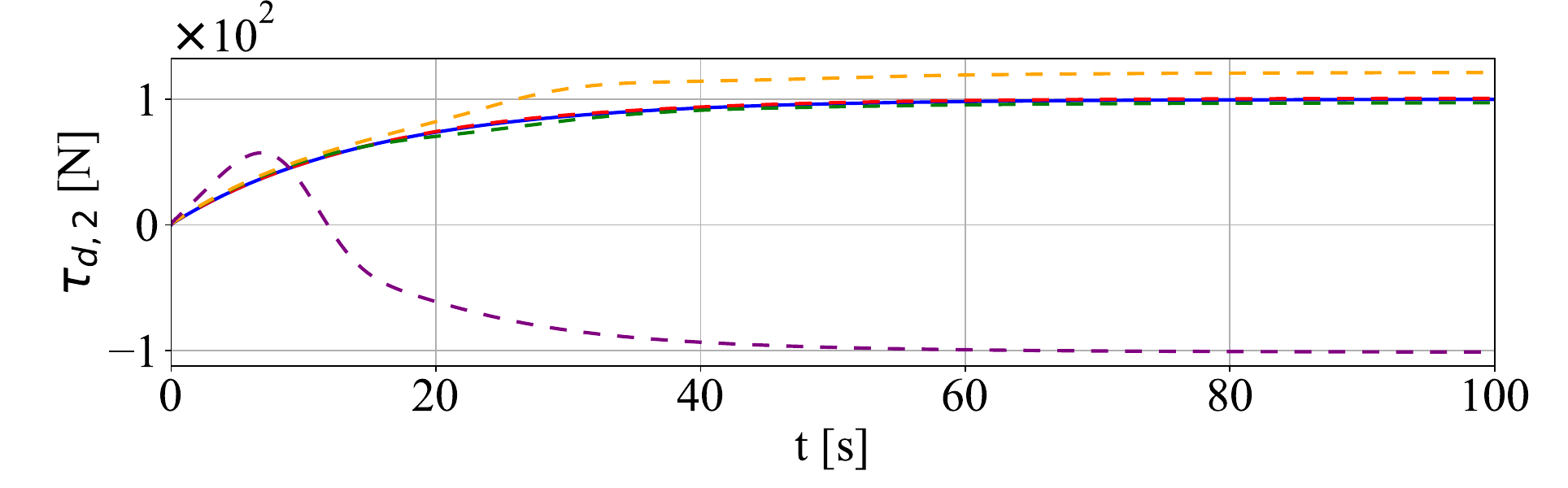}
		\caption{Observed disturbances concerning sway}
	\end{subfigure}
	\begin{subfigure}{0.33\linewidth}
		\includegraphics[width=\linewidth]{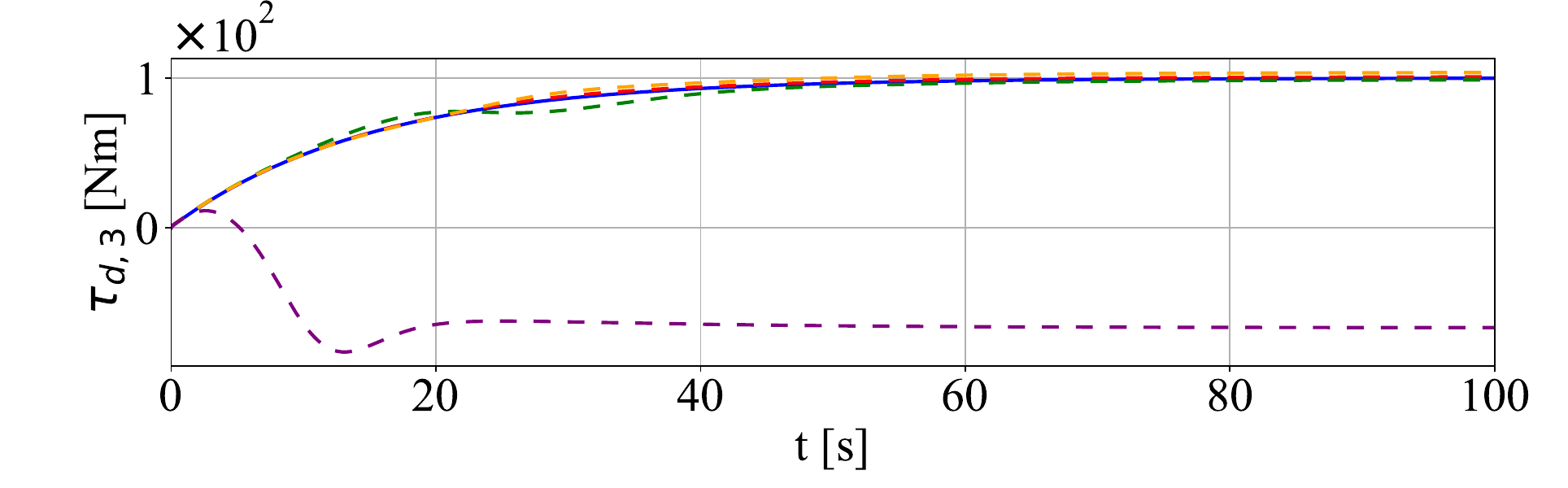} 
		\caption{Observed disturbances concerning rotational speed}
	\end{subfigure}
	\caption{Observed disturbances under consideration of various randomly generated model uncertainties, where the model parameters are created with uncertainties $\rho=\rho_1$ (\protect\redlinedashed), $\rho=\rho_2$ (\protect\greenlinedashed), $\rho=\rho_3$ (\protect\orangelinedashed), $\rho=\rho_4$ (\protect\purplelinedashed), and the results are compared to the actual disturbances (\protect\blueline).}
	\label{fig: Model_uncertainties}
\end{figure*}

In \figref{fig: Model_uncertainties}, the estimations of the observed disturbances are showcased if the model parameters are wrongly identified. If the model parameters have an uncertainty below 10\%, the estimations are still relatively accurate. In this case study, the observer starts getting inaccurate if the model uncertainties exceed 10\%, and the estimations are useless if the identified model parameters are intensively erroneous (30\%).

Since the model is described in dependency of the system states, and these system states are considered to be measured, the model and measurements are correlated.
Hence, the influence of both model and measurement uncertainties leads to a superposition. As a result, the framework can deal with them in parallel until the model uncertainties exceed a specific limit depicted in \figref{fig: Model_uncertainties}. These scenarios are separated to avoid confusion and clearly state the individual capabilities.


\section{Conclusion and future work}
\label{sec:conclusionandfuturework}
The proposed disturbance observer framework proved its capability for approximating the environmental impact of wind, waves, and sea currents despite using an unreliable model and highly noisy measurements. 
The main takeaway from the current work can be itemized as follows:
\begin{itemize}	
	\item The cascaded structure of the observer framework allows adaptive smoothing of noisy measurements. As a result, the observed disturbances have approximately no noisy characteristics. 
	\item Model simplifications are avoided, and the observer framework showed that the disturbances can be approximated relatively accurately despite model uncertainties. However, the observation capabilities are limited if the model parameters are highly erroneously identified.
	\item Under consideration of a discrete, temporal sequence of measurements, the framework can observe the actual disturbances despite low sampling rates. 
	\item The formulation of the observer's error dynamics allows the user a convenient approach for synchronizing the adaptive gains. As a result, the framework can adapt its estimations with a synchronized adaptation speed and equal smoothing properties.
\end{itemize}
Future work will utilize the proposed disturbance observer framework for robust control of autonomous surface vessels and concentrate on improving their situational awareness, e.g., via online parameter identification, to overcome the issue of false estimations by using erroneous model parameters and guaranteeing trustworthy model dynamics.


\section*{Acknowledgments}
This work is part of SFI AutoShip, an 8-year research-based innovation center. 
In addition, this research project is integrated into the PERSEUS doctoral program. 
We want to thank our partners, including the Research Council of Norway, under project number 309230, and the European Union’s Horizon 2020 research and innovation program under the Marie Skłodowska-Curie grant agreement number 101034240.

\bibliographystyle{AR} 
\bibliography{manuscript}

\begin{thebibliography}{10}
\expandafter\ifx\csname url\endcsname\relax
  \def\url#1{\texttt{#1}}\fi
\expandafter\ifx\csname urlprefix\endcsname\relax\def\urlprefix{URL }\fi
\expandafter\ifx\csname href\endcsname\relax
  \def\href#1#2{#2} \def\path#1{#1}\fi

\bibitem{Do2005}
K.~Do, Z.~Jiang, J.~Pan,
  \href{https://dx.doi.org/10.1016/j.sysconle.2005.02.014}{Global partial-state
  feedback and output-feedback tracking controllers for underactuated ships},
  {\em Systems {\&} Control Letters}, 54:1015--1036 (2005).
\newblock

\bibitem{Wei2022}
X.~Wei, L.~You, H.~Zhang, X.~Hu, J.~Han,
  \href{https://dx.doi.org/10.1002/rnc.6023}{Disturbance observer based control
  for dynamically positioned ships with ocean environmental disturbances and
  actuator saturation}, {\em International Journal of Robust and Nonlinear
  Control}, 32:4113--4128 (2022).
\newblock

\bibitem{Huang2015}
J.~Huang, C.~Wen, W.~Wang, Y.-D. Song,
  \href{https://dx.doi.org/10.1016/j.sysconle.2015.07.002}{Global stable
  tracking control of underactuated ships with input saturation}, {\em Systems
  {\&} Control Letters}, 85:1--7 (2015).
\newblock

\bibitem{Hu2022}
X.~Hu, X.~Wei, Y.~Kao, J.~Han,
  \href{https://dx.doi.org/10.1109/tits.2021.3054177}{{Robust Synchronization
  for Under-Actuated Vessels Based on Disturbance Observer}}, {\em {IEEE}
  Transactions on Intelligent Transportation Systems}, 23:5470--5479 (2022).
\newblock

\bibitem{Peng2021a}
Z.~Peng, J.~Wang, D.~Wang, Q.-L. Han,
  \href{https://dx.doi.org/10.1109/tii.2020.3004343}{{An Overview of Recent
  Advances in Coordinated Control of Multiple Autonomous Surface Vehicles}},
  {\em {IEEE} Transactions on Industrial Informatics}, 17:732--745 (2021).
\newblock

\bibitem{Li2015}
Y.~Li, B.~Yang, T.~Zheng, Y.~Li, M.~Cui, S.~Peeta,
  \href{https://dx.doi.org/10.1109/tits.2015.2410282}{{Extended-State-Observer-Based
  Double-Loop Integral Sliding-Mode Control of Electronic Throttle Valve}},
  {\em {IEEE} Transactions on Intelligent Transportation Systems},
  16:2501--2510 (2015).
\newblock

\bibitem{Chen2019}
Z.~Chen, Y.~Zhang, Y.~Zhang, Y.~Nie, J.~Tang, S.~Zhu,
  \href{https://dx.doi.org/10.1109/access.2019.2941364}{{Disturbance-Observer-Based
  Sliding Mode Control Design for Nonlinear Unmanned Surface Vessel With
  Uncertainties}}, {\em {IEEE} Access}, 7:148522--148530 (2019).
\newblock

\bibitem{Selvaraj2020}
P.~Selvaraj, O.~Kwon, S.~Lee, R.~Sakthivel,
  \href{https://dx.doi.org/10.1016/j.jfranklin.2020.04.016}{Uncertainty and
  disturbance rejections of complex dynamical networks via truncated predictive
  control}, {\em Journal of the Franklin Institute}, 357:4901--4921 (2020).
\newblock

\bibitem{Hu2019}
X.~Hu, X.~Wei, H.~Zhang, J.~Han, X.~Liu,
  \href{https://dx.doi.org/10.1007/s11071-018-4692-1}{Global asymptotic
  regulation control for {MIMO} mechanical systems with unknown model
  parameters and disturbances}, {\em Nonlinear Dynamics}, 95:2293--2305 (2019).
\newblock

\bibitem{Bouteraa2022}
Y.~Bouteraa, K.~A. Alattas, S.~Mobayen, M.~Golestani, A.~Ibrahim, U.~Tariq,
  \href{https://dx.doi.org/10.3390/act11050128}{{Disturbance Observer-Based
  Tracking Controller for Uncertain Marine Surface Vessel}}, {\em Actuators},
  11:128 (2022).
\newblock

\bibitem{Duong2022}
N.~T. Duong, N.~Q. Duy,
  \href{https://dx.doi.org/10.11591/ijece.v12i2.pp1392-1401}{Adaptive
  backstepping control for ship nonlinear active fin system based on
  disturbance observer and neural network}, {\em International Journal of
  Electrical and Computer Engineering ({IJECE})}, 12:1392 (2022).
\newblock

\bibitem{Xu2022}
D.~Xu, Z.~Liu, X.~Zhou, L.~Yang, L.~Huang,
  \href{https://dx.doi.org/10.3390/app12063004}{{Trajectory Tracking of
  Underactuated Unmanned Surface Vessels: Non-Singular Terminal Sliding Control
  with Nonlinear Disturbance Observer}}, {\em Applied Sciences}, 12:3004
  (2022).
\newblock

\bibitem{Do2010}
K.~D. Do, \href{https://dx.doi.org/10.1016/j.oceaneng.2010.04.007}{Practical
  control of underactuated ships}, {\em Ocean Engineering}, 37:1111--1119
  (2010).
\newblock

\bibitem{Gu2022}
N.~Gu, D.~Wang, Z.~Peng, J.~Wang, Q.-L. Han,
  \href{https://dx.doi.org/10.1016/j.conengprac.2022.105158}{{Disturbance
  observers and extended state observers for marine vehicles: A survey}}, {\em
  Control Engineering Practice}, 123:105158 (2022).
\newblock

\bibitem{Huang2019}
C.~Huang, X.~Zhang, Y.~Deng, G.~Zhang, {Robust Dynamic Positioning Control of
  Marine Ships via a Disturbance Observer}, {\em Proceedings of the
  Twenty-ninth (2019) International Ocean and Polar Engineering Conference}
  (2019).

\bibitem{Qiang2019}
Z.~Qiang, Z.~Guibing, H.~Xin, Y.~Renming,
  \href{https://dx.doi.org/10.1016/j.oceaneng.2019.02.031}{Adaptive neural
  network auto-berthing control of marine ships}, {\em Ocean Engineering},
  177:40--48 (2019).
\newblock

\bibitem{Peng2021}
Z.~Peng, D.~Wang, J.~Wang,
  \href{https://dx.doi.org/10.1109/tnnls.2021.3093330}{{Data-Driven Adaptive
  Disturbance Observers for Model-Free Trajectory Tracking Control of Maritime
  Autonomous Surface Ships}}, {\em {IEEE} Transactions on Neural Networks and
  Learning Systems}, 32:5584--5594 (2021).
\newblock

\bibitem{Trivedi2021}
U.~B. Trivedi, M.~Bhatt, P.~Srivastava,
  \href{https://dx.doi.org/10.1007/978-3-030-66218-9\_40}{{Prevent Overfitting
  Problem in Machine Learning: A Case Focus on Linear Regression and Logistics
  Regression}}, in {\em Innovations in Information and Communication
  Technologies ({IICT}-2020)}, Springer International Publishing, 2021, pp.
  345--349.
\newblock

\bibitem{Emami2019}
S.~A. Emami, A.~Banazadeh,
  \href{https://dx.doi.org/10.1002/rnc.4698}{Intelligent trajectory tracking of
  an aircraft in the presence of internal and external disturbances}, {\em
  International Journal of Robust and Nonlinear Control}, 29:5820--5844 (2019).
\newblock

\bibitem{Chen2019a}
G.~Chen, S.~Chen, R.~Langari, X.~Li, W.~Zhang,
  \href{https://dx.doi.org/10.1109/tvt.2019.2922452}{{Driver-Behavior-Based
  Adaptive Steering Robust Nonlinear Control of Unmanned Driving Robotic
  Vehicle With Modeling Uncertainties and Disturbance Observer}}, {\em {IEEE}
  Transactions on Vehicular Technology}, 68:8183--8190 (2019).
\newblock

\bibitem{Kim2003}
M.-S. Kim, J.-H. Shin, S.-G. Hong, J.-J. Lee,
  \href{https://dx.doi.org/10.1016/s0957-4158(02)00002-8}{Designing a robust
  adaptive dynamic controller for nonholonomic mobile robots under modeling
  uncertainty and disturbances}, {\em Mechatronics}, 13:507--519 (2003).
\newblock

\bibitem{Fossen2011}
T.~I. Fossen, \href{https://dx.doi.org/10.1002/9781119994138}{{\em {Handbook of
  Marine Craft Hydrodynamics and Motion Control}}}, John Wiley {\&} Sons, Ltd,
  2011.
\newblock

\bibitem{Merigo2019}
J.~M. Merig{\'{o}}, R.~R. Yager,
  \href{https://dx.doi.org/10.1007/s00500-019-03892-w}{Aggregation operators
  with moving averages}, {\em Soft Computing}, 23:10601--10615 (2019).
\newblock

\bibitem{Yager2008}
R.~R. Yager, \href{https://dx.doi.org/10.1109/tfuzz.2008.917299}{{Time Series
  Smoothing and {OWA} Aggregation}}, {\em {IEEE} Transactions on Fuzzy
  Systems}, 16:994--1007 (2008).
\newblock

\bibitem{Julier1997}
S.~J. Julier, J.~K. Uhlmann, \href{https://dx.doi.org/10.1117/12.280797}{{New
  extension of the Kalman filter to nonlinear systems}}, in I.~Kadar (ed.),
  {\em {SPIE} Proceedings}, {SPIE}, 1997.
\newblock

\bibitem{Brekke2022}
E.~F. Brekke, E.~Eide, B.-O.~H. Eriksen, E.~F. Wilthil, M.~Breivik,
  E.~Skjellaug, Øystein K.~Helgesen, A.~M. Lekkas, A.~B. Martinsen, E.~H.
  Thyri,
  \href{https://dx.doi.org/10.1088/1742-6596/2311/1/012029}{{milliAmpere: An
  Autonomous Ferry Prototype}}, {\em Journal of Physics: Conference Series},
  2311:012029 (2022).
\newblock

\bibitem{Pedersen2019}
A.~A. Pedersen, {Optimization Based System Identification for the
  milliAmpereFerry}, Master's thesis, Norwegian University of Science and
  Technology (2019).

\end{thebibliography}

\section*{Appendix}
\begin{table}[H]
	\centering
	\caption{Parameters of the observer framework}
	\label{tab:Initialization}
	\begin{tabular}{ll}
		\toprule[1.5pt]
		\bf Parameter & \bf Value \\
		\midrule
		\midrule
		$\mathrm{Initializations}$\\
		\midrule
		$\boldsymbol{\hat{\nu}}_0$ & $\boldsymbol{E}[\boldsymbol{\nu}_0]$ \\
		$\boldsymbol{P}_0$ & $0.1\cdot\mathrm{eye}(L,L)$ \\
		$\boldsymbol{\zeta}_{0,j}$ & $\mathrm{zeros}(L,1)$ \\
		\midrule
		$\mathrm{Observer \ parameters}$\\
		\midrule
		$\Gamma_{i,1}$ & $15$ \\
		$\Gamma_{i,2}$ & $3$ \\
		$\Gamma_{i,3}$ & $0.2$ \\
		\midrule
		$\mathrm{UKF \ parameters}$\\
		\midrule
		$\alpha$ & $10^{-3}$ \\
		$\beta$ & $2$ \\
		$\kappa$ & $0$ \\
		\midrule
		$\mathrm{Noise \ estimator \ parameters}$\\
		\midrule
		$w$ & $100$ \\
		$\gamma_1$ & $0.001$ \\
		$\gamma_2$ & $0.005$ \\
		$\gamma_3$ & $0.01$ \\
		\bottomrule
	\end{tabular}
\end{table}

\begin{table}[H]
	\centering
	\caption{Identified parameters of the milliAmpere ferry}
	\label{tab: Parameters}
	\begin{tabular}{lll | lll}
		\toprule[1.5pt]
		{\bf Par} & {\bf Value} & {\bf Unit} & {\bf Par} & {\bf Value} & {\bf Unit}\\
		\midrule
		$m_{11}$ & 2389.657 &\unit{kg}  & $m_{22}$ & 2533.911 &\unit{kg} \\
		$m_{23}$ & 62.386 &\unit{$\mathrm{kg}$}$\cdot$\unit{$\mathrm{m}$}    & $m_{32}$ & 28.141 &\unit{$\mathrm{kg}$}$\cdot$\unit{$\mathrm{m}$} \\
		$m_{33}$ & 5068.910 &\unit{$\mathrm{kg}$}$\cdot$\unit{$\mathrm{m^2}$} & $X_u$ & -27.632 &\unit{$\mathrm{kg}$}$\cdot$\unit{$\mathrm{s^{-1}}$}\\
		$X_{|u|u}$& -110.064 &\unit{$\mathrm{kg}$}$\cdot$\unit{$\mathrm{m^{-1}}$} & $X_{uuu}$& -13.965 &\unit{$\mathrm{kg}$}$\cdot$\unit{$\mathrm{s}$}$\cdot$\unit{$\mathrm{m^{-1}}$} \\
		$Y_v$& -52.947 &\unit{$\mathrm{kg}$}$\cdot$\unit{$\mathrm{s^{-1}}$} &	$Y_{|v|v}$ & -116.486 &\unit{$\mathrm{kg}$}$\cdot$\unit{$\mathrm{m^{-1}}$}\\
		$Y_{vvv}$& -24.313 &\unit{$\mathrm{kg}$}$\cdot$\unit{$\mathrm{s}$}$\cdot$\unit{$\mathrm{m^{-1}}$} &	$Y_{|r|v}$& -1540.383 &\unit{$\mathrm{kg}$}\\
		$Y_r$& 24.732 &\unit{$\mathrm{kg}$}$\cdot$\unit{$\mathrm{m}$}$\cdot$\unit{$\mathrm{s^{-1}}$} &	$Y_{|v|r}$& 572.141 &\unit{$\mathrm{kg}$}\\
		$Y_{|r|r}$& -115.457 &\unit{$\mathrm{kg}$}$\cdot$\unit{$\mathrm{m}$} & $N_{v}$& 3.5241 &\unit{$\mathrm{kg}$}$\cdot$\unit{$\mathrm{m}$}$\cdot$\unit{$\mathrm{s^{-1}}$}\\
		$N_{|v|v}$& -0.832 &\unit{$\mathrm{kg}$} & $N_{|r|v}$& 336.827 &\unit{$\mathrm{kg}$}$\cdot$\unit{$\mathrm{m}$}\\
		$N_{r}$& -122.860 &\unit{$\mathrm{kg}$}$\cdot$\unit{$\mathrm{m^2}$}$\cdot$\unit{$\mathrm{s^{-1}}$}& $N_{|r|r}$& -874.428 &\unit{$\mathrm{kg}$}$\cdot$\unit{$\mathrm{m^2}$}\\
		$N_{rrr}$& 0.000 &\unit{$\mathrm{kg}$}$\cdot$\unit{$\mathrm{m^2}$}$\cdot$\unit{$\mathrm{s}$} & $N_{|v|r}$& -121.957 &\unit{$\mathrm{kg}$}$\cdot$\unit{$\mathrm{m}$}\\
		\bottomrule
	\end{tabular}
\end{table}

\begin{table}[H]
	\centering
	\caption{Simulation parameters}
	\label{tab: Simulation parameters}
	\begin{tabular}{lllll}
		\toprule[1.5pt]
		{\bf Par} & {\bf Value} & {\bf Unit}\\
		\midrule
		\midrule
		$\mathrm{Initializations}$\\
		\midrule
		$\boldsymbol{\eta}$ & $[0,0,30]^\top$ & \unit{[m,m,deg]}\\
		$\boldsymbol{\nu}$ & $[0,0,0]^\top$ & [$\frac{\unit{m}}{\unit{s}}$,$\frac{\unit{m}}{\unit{s}}$,$\frac{\unit{deg}}{\unit{s}}$]\\
		\midrule
		$\mathrm{Simulation \ parameters}$\\
		\midrule		
		$\Delta t$ & 0.01 & \unit{s}\\
		$\bar{F}_{\mathrm{wind}}$ & 50 & \unit{N}\\
		$\bar{F}_{\mathrm{wave}}$ & 20 & \unit{N}\\
		$\bar{F}_{\mathrm{current}}$ & 100 & \unit{N}\\
		$\gamma_{\mathrm{wind}}$ & 135 & \unit{deg}\\
		$\gamma_{\mathrm{wave}}$ & 155 & \unit{deg}\\
		$\gamma_{\mathrm{current}}$ & 270 & \unit{deg}\\
		$L$ & 5 & \unit{m}\\
		$T_s$ & 15 & \unit{s}\\
		\midrule
		$\mathrm{Uncertainty \ parameters}$\\
		\midrule
		$\boldsymbol{R}_1$ & $2\cdot10^{-3}\boldsymbol{I}_{3\times3}$\\
		$\boldsymbol{R}_2$ & $5\cdot10^{-3}\boldsymbol{I}_{3\times3}$\\
		$\boldsymbol{R}_3$ & $10^{-2}\boldsymbol{I}_{3\times3}$\\
		$\boldsymbol{R}_4$ & $10^{-1}\boldsymbol{I}_{3\times3}$\\
		$\rho_1$ & $0.01$\\
		$\rho_2$ & $0.05$\\
		$\rho_3$ & $0.1$\\
		$\rho_4$ & $0.3$\\
		\bottomrule
	\end{tabular}
\end{table}

\end{document}